\documentclass[superscriptaddress, amsmath,amssymb, aps,twocolumn,longbibliography]{revtex4-1} 
\usepackage{graphicx}
\usepackage{dcolumn}
\usepackage{amsthm}
\usepackage{bm}

\pdfoutput=1
\usepackage{graphicx,graphics,epsfig,subfigure,times,bm,bbm,amssymb,amsmath,amsthm,mathrsfs,MnSymbol}
\usepackage{gensymb}
\usepackage{amsfonts}
\usepackage[pdfstartview=FitH]{hyperref}
\hypersetup{
    colorlinks=true,       
    linkcolor=red,          
    citecolor=magenta,        
    filecolor=magenta,      
    urlcolor=cyan,           
    runcolor=cyan}
\usepackage{epstopdf}
\usepackage[pdftex]{color}
\usepackage{braket} 
\usepackage{enumerate}
\usepackage[normalem]{ulem}
\usepackage[usenames,dvipsnames]{xcolor}
\usepackage{multirow}
\usepackage{mathtools}
\usepackage{relsize} 

\definecolor{orange}{rgb}{1,0.5,0}

\usepackage{changes}

\begin{document}

\title{Many-body delocalization with a two-dimensional 70-qubit superconducting quantum simulator}

\author{Tian-Ming Li}
    \thanks{These authors contributed equally to this work.}
    \affiliation{Beijing National Laboratory for Condensed Matter Physics, Institute of Physics, Chinese Academy of Sciences, Beijing 100190, China}
    \affiliation{School of Physical Sciences, University of Chinese Academy of Sciences, Beijing 100049, China}

\author{Zheng-Hang Sun}
    \thanks{These authors contributed equally to this work.}
    \affiliation{Theoretical Physics \uppercase\expandafter{\romannumeral3}, Center for Electronic Correlations and Magnetism, Institute of Physics, University of Augsburg, D-86135 Augsburg, Germany}

\author{Yun-Hao Shi}
    \thanks{These authors contributed equally to this work.}
    \affiliation{Beijing National Laboratory for Condensed Matter Physics, Institute of Physics, Chinese Academy of Sciences, Beijing 100190, China}

\author{Zhen-Ting Bao}
    \thanks{These authors contributed equally to this work.}
    \affiliation{Beijing National Laboratory for Condensed Matter Physics, Institute of Physics, Chinese Academy of Sciences, Beijing 100190, China}
    \affiliation{School of Physical Sciences, University of Chinese Academy of Sciences, Beijing 100049, China}
    
\author{Yong-Yi Wang}
    \affiliation{Beijing National Laboratory for Condensed Matter Physics, Institute of Physics, Chinese Academy of Sciences, Beijing 100190, China}
    \affiliation{School of Physical Sciences, University of Chinese Academy of Sciences, Beijing 100049, China}

\author{Jia-Chi Zhang}
    \affiliation{Beijing National Laboratory for Condensed Matter Physics, Institute of Physics, Chinese Academy of Sciences, Beijing 100190, China}
    \affiliation{School of Physical Sciences, University of Chinese Academy of Sciences, Beijing 100049, China}

\author{Yu Liu}
    \affiliation{Beijing National Laboratory for Condensed Matter Physics, Institute of Physics, Chinese Academy of Sciences, Beijing 100190, China}
    \affiliation{School of Physical Sciences, University of Chinese Academy of Sciences, Beijing 100049, China}
    
\author{Cheng-Lin Deng}
    \affiliation{Beijing Key Laboratory of Fault-Tolerant Quantum Computing, Beijing Academy of Quantum Information Sciences, Beijing 100193, China}

\author{Yi-Han Yu}
    \affiliation{Beijing National Laboratory for Condensed Matter Physics, Institute of Physics, Chinese Academy of Sciences, Beijing 100190, China}
    \affiliation{School of Physical Sciences, University of Chinese Academy of Sciences, Beijing 100049, China}

\author{Zheng-He Liu}
    \affiliation{Beijing National Laboratory for Condensed Matter Physics, Institute of Physics, Chinese Academy of Sciences, Beijing 100190, China}
    \affiliation{School of Physical Sciences, University of Chinese Academy of Sciences, Beijing 100049, China}

\author{Chi-Tong Chen}
    \affiliation{Quantum Science Center of Guangdong-Hong Kong-Macao Greater Bay Area, Shenzhen, Guangdong 518045, China}

\author{Li Li}
    \affiliation{Beijing National Laboratory for Condensed Matter Physics, Institute of Physics, Chinese Academy of Sciences, Beijing 100190, China}
    \affiliation{School of Physical Sciences, University of Chinese Academy of Sciences, Beijing 100049, China}

\author{Hao Li}
    \affiliation{Beijing Key Laboratory of Fault-Tolerant Quantum Computing, Beijing Academy of Quantum Information Sciences, Beijing 100193, China}

\author{Hao-Tian Liu}
    \affiliation{Beijing National Laboratory for Condensed Matter Physics, Institute of Physics, Chinese Academy of Sciences, Beijing 100190, China}
    \affiliation{School of Physical Sciences, University of Chinese Academy of Sciences, Beijing 100049, China}
    \affiliation{Beijing Key Laboratory of Fault-Tolerant Quantum Computing, Beijing Academy of Quantum Information Sciences, Beijing 100193, China}

\author{Si-Yun Zhou}
    \affiliation{Beijing National Laboratory for Condensed Matter Physics, Institute of Physics, Chinese Academy of Sciences, Beijing 100190, China}
    \affiliation{School of Physical Sciences, University of Chinese Academy of Sciences, Beijing 100049, China}
    
\author{Zhen-Yu Peng}
    \affiliation{Beijing National Laboratory for Condensed Matter Physics, Institute of Physics, Chinese Academy of Sciences, Beijing 100190, China}
    \affiliation{School of Physical Sciences, University of Chinese Academy of Sciences, Beijing 100049, China}

\author{Yan-Jun Liu}
    \affiliation{Beijing National Laboratory for Condensed Matter Physics, Institute of Physics, Chinese Academy of Sciences, Beijing 100190, China}
    \affiliation{School of Physical Sciences, University of Chinese Academy of Sciences, Beijing 100049, China}

\author{Ziting Wang}
    \affiliation{Beijing Key Laboratory of Fault-Tolerant Quantum Computing, Beijing Academy of Quantum Information Sciences, Beijing 100193, China}

\author{Yueshan Xu}
    \affiliation{Beijing Key Laboratory of Fault-Tolerant Quantum Computing, Beijing Academy of Quantum Information Sciences, Beijing 100193, China}

\author{Kui Zhao}
    \affiliation{Beijing Key Laboratory of Fault-Tolerant Quantum Computing, Beijing Academy of Quantum Information Sciences, Beijing 100193, China}

\author{Yang He}
    \affiliation{Beijing National Laboratory for Condensed Matter Physics, Institute of Physics, Chinese Academy of Sciences, Beijing 100190, China}
    \affiliation{School of Physical Sciences, University of Chinese Academy of Sciences, Beijing 100049, China}

\author{Da'er Feng}
    \affiliation{Beijing National Laboratory for Condensed Matter Physics, Institute of Physics, Chinese Academy of Sciences, Beijing 100190, China}
    \affiliation{School of Physical Sciences, University of Chinese Academy of Sciences, Beijing 100049, China}

\author{Jia-Cheng Song}
    \affiliation{Beijing National Laboratory for Condensed Matter Physics, Institute of Physics, Chinese Academy of Sciences, Beijing 100190, China}
    \affiliation{School of Physical Sciences, University of Chinese Academy of Sciences, Beijing 100049, China}

\author{Cai-Ping Fang}
    \affiliation{Beijing National Laboratory for Condensed Matter Physics, Institute of Physics, Chinese Academy of Sciences, Beijing 100190, China}
    \affiliation{School of Physical Sciences, University of Chinese Academy of Sciences, Beijing 100049, China}
    \affiliation{Beijing Key Laboratory of Fault-Tolerant Quantum Computing, Beijing Academy of Quantum Information Sciences, Beijing 100193, China}

\author{Junrui Deng}
    \affiliation{Beijing National Laboratory for Condensed Matter Physics, Institute of Physics, Chinese Academy of Sciences, Beijing 100190, China}
    \affiliation{School of Physical Sciences, University of Chinese Academy of Sciences, Beijing 100049, China}

\author{Mingyu Xu}
    \affiliation{Beijing National Laboratory for Condensed Matter Physics, Institute of Physics, Chinese Academy of Sciences, Beijing 100190, China}
    \affiliation{School of Physical Sciences, University of Chinese Academy of Sciences, Beijing 100049, China}

\author{Yu-Tao Chen}
    \affiliation{Beijing National Laboratory for Condensed Matter Physics, Institute of Physics, Chinese Academy of Sciences, Beijing 100190, China}
    \affiliation{School of Physical Sciences, University of Chinese Academy of Sciences, Beijing 100049, China}

\author{Bozhen Zhou}
    \affiliation{Institute of Theoretical Physics, Chinese Academy of Sciences, Beijing 100190, China}

\author{Gui-Han Liang}
    \affiliation{Beijing National Laboratory for Condensed Matter Physics, Institute of Physics, Chinese Academy of Sciences, Beijing 100190, China}
    \affiliation{School of Physical Sciences, University of Chinese Academy of Sciences, Beijing 100049, China}

\author{Zhongcheng Xiang}
    \affiliation{Beijing National Laboratory for Condensed Matter Physics, Institute of Physics, Chinese Academy of Sciences, Beijing 100190, China}
    \affiliation{School of Physical Sciences, University of Chinese Academy of Sciences, Beijing 100049, China}

\author{Guangming Xue}
    \affiliation{Beijing Key Laboratory of Fault-Tolerant Quantum Computing, Beijing Academy of Quantum Information Sciences, Beijing 100193, China}

\author{Dongning Zheng}
    \affiliation{Beijing National Laboratory for Condensed Matter Physics, Institute of Physics, Chinese Academy of Sciences, Beijing 100190, China}
    \affiliation{School of Physical Sciences, University of Chinese Academy of Sciences, Beijing 100049, China}

\author{Kaixuan Huang}
    \affiliation{Beijing Key Laboratory of Fault-Tolerant Quantum Computing, Beijing Academy of Quantum Information Sciences, Beijing 100193, China}

\author{Zheng-An Wang}
    \affiliation{Beijing Key Laboratory of Fault-Tolerant Quantum Computing, Beijing Academy of Quantum Information Sciences, Beijing 100193, China}    
    
\author{Haifeng Yu}
    \email{hfyu@baqis.ac.cn}
    \affiliation{Beijing Key Laboratory of Fault-Tolerant Quantum Computing, Beijing Academy of Quantum Information Sciences, Beijing 100193, China}

\author{Piotr Sierant}
    \affiliation{Barcelona Supercomputing Center, Barcelona 08034, Spain}
    
\author{Kai Xu}
    \email{kaixu@iphy.ac.cn}
    \affiliation{Beijing National Laboratory for Condensed Matter Physics, Institute of Physics, Chinese Academy of Sciences, Beijing 100190, China}
    \affiliation{School of Physical Sciences, University of Chinese Academy of Sciences, Beijing 100049, China}
    \affiliation{Beijing Key Laboratory of Fault-Tolerant Quantum Computing, Beijing Academy of Quantum Information Sciences, Beijing 100193, China}
    \affiliation{Hefei National Laboratory, Hefei 230088, China}
    \affiliation{Songshan Lake Materials Laboratory, Dongguan, Guangdong 523808, China}

\author{Heng Fan}
    \email{hfan@iphy.ac.cn}
    \affiliation{Beijing National Laboratory for Condensed Matter Physics, Institute of Physics, Chinese Academy of Sciences, Beijing 100190, China}
    \affiliation{School of Physical Sciences, University of Chinese Academy of Sciences, Beijing 100049, China}
    \affiliation{Beijing Key Laboratory of Fault-Tolerant Quantum Computing, Beijing Academy of Quantum Information Sciences, Beijing 100193, China}
    \affiliation{Hefei National Laboratory, Hefei 230088, China}
    \affiliation{Songshan Lake Materials Laboratory, Dongguan, Guangdong 523808, China}

\begin{abstract}
Quantum many-body systems with sufficiently strong disorder can exhibit a non-equilibrium phenomenon, known as the many-body localization (MBL), which is distinct from conventional thermalization. 
While the MBL regime has been extensively studied in one dimension, its existence in higher dimensions remains elusive, challenged by the avalanche instability. Here, using a 70-qubit two-dimensional (2D) superconducting quantum simulator, we experimentally explore the robustness of the MBL regime in controlled finite-size 2D systems. We observe that the decay of imbalance becomes more pronounced with increasing system sizes, scaling up from 21, 42 to 70 qubits, with a relatively large disorder strength, and for the first time, provide an evidence for the many-body delocalization in 2D disordered systems. Our experimental results are consistent with the avalanche theory that predicts the instability of MBL regime beyond one spatial dimension. This work establishes a scalable platform for probing high-dimensional non-equilibrium phases of matter and their finite-size effects using superconducting quantum circuits.
\end{abstract}
\maketitle

Understanding how quantum particles become localized and exhibit non-ergodic behavior beyond the expectations of the eigenstate thermalization hypothesis (ETH) is a central topic in the field of quantum statistical mechanics~\cite{Eisert:2015ws}. The phenomenon of Anderson localization demonstrates that random disorder can localize non-interacting quantum particles~\cite{PhysRev.109.1492}. 
In the presence of strong interactions, it has been proposed that sufficiently strong disorder can give rise to a many-body localization (MBL) regime, in which the ETH is violated even at long (but finite) evolution times and in large (but finite) system sizes~\cite{annurev:/content/journals/10.1146/annurev-conmatphys-031214-014726,Altman:2018vb,RevModPhys.91.021001}. In recent years, considerable effort, particularly in one-dimensional (1D) systems, has been devoted to understanding the conditions under which the MBL regime gives rise to a genuine MBL phase~\cite{Sierant_2025}, in which non-ergodic behavior persists even in the limit of \emph{infinite} time and system size. Finite-size scaling analysis based on the Berezinskii–Kosterlitz–Thouless correlation length suggests that the ETH-MBL crossover point tends to infinity in the thermodynamic limit~\cite{PhysRevE.102.062144,PhysRevB.102.064207,PhysRevB.111.094210}, whereas the avalanche theory indicates that an MBL phase can occur much deeper in the finite-size MBL regimes~\cite{PhysRevB.105.174205,PhysRevB.106.L020202,PhysRevB.108.L020201,PhysRevB.109.134202}. For 1D systems, it is now recognized that the Poisson level statistics, as a conventional landmark, 
underestimates the disorder strength required for the stable MBL phase~\cite{PhysRevLett.125.156601, PhysRevB.105.174205}.

\begin{figure*}[]
	\centering
	\includegraphics[width=1\linewidth]{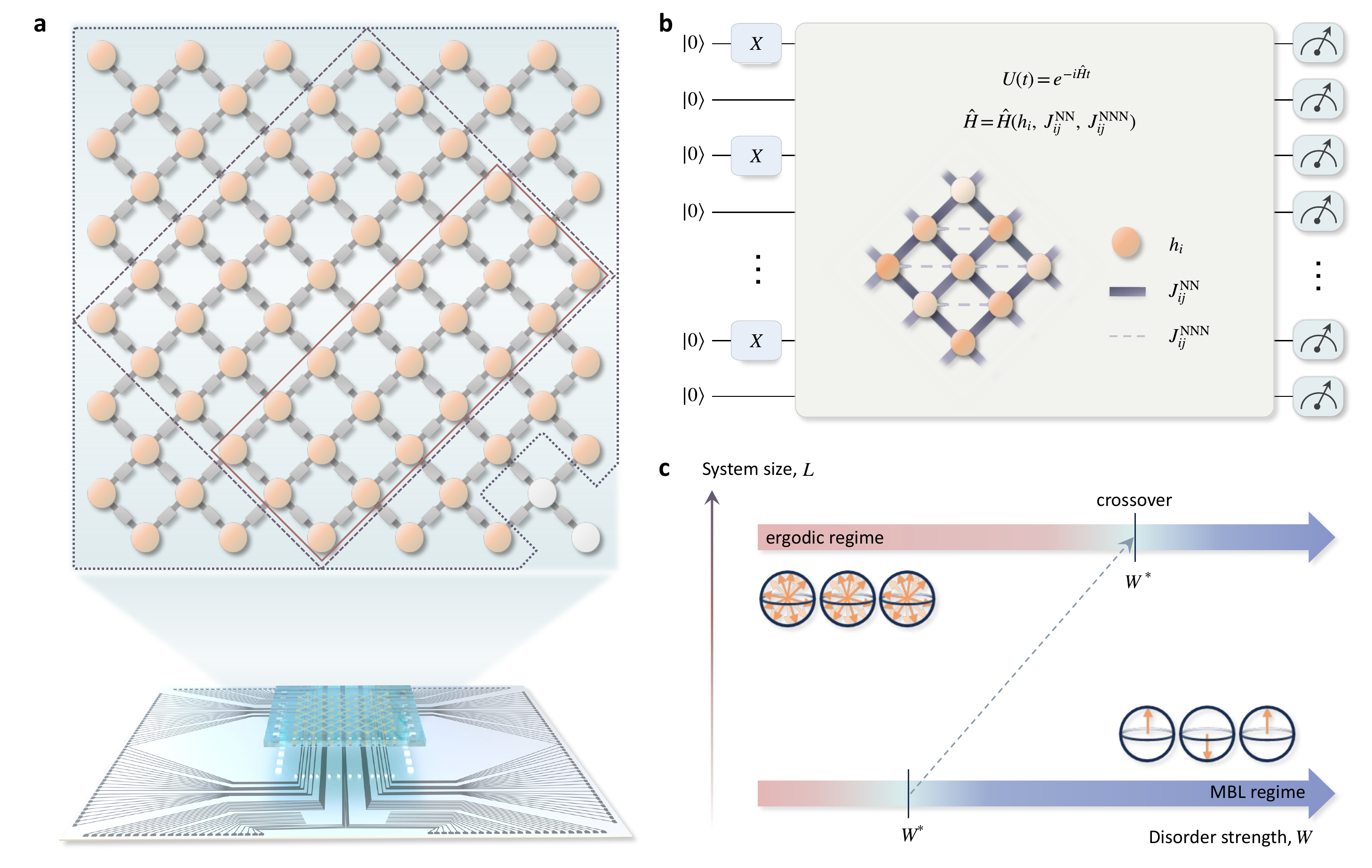}\\
	\caption{\textbf{Superconducting quantum simulator and experimental scheme.} \textbf{a,}  Schematic of the 2D superconducting quantum simulator, which consists of 70 qubits and interconnected by 117 couplers linking all adjacent qubits. The red solid-line rectangle, the black dashed-line diamond and the gray dotted-line square highlight the selected qubits for the system sizes $L=21$, $42$ and $70$, respectively. \textbf{b,} Schematic representation of the analog quantum simulation experiment. The half-filling initial state is prepared by implementing $X$ gate on $n=L/2$ selected qubits. The system is then evolved under the Hamiltonian (\ref{hamiltonian}), with the coupling strengths $J_{ij}^\mathsmaller{\rm NN}$, $J_{ij}^\mathsmaller{\rm NNN}$, and tunable disorder $h_{i}$, and finally the qubits are measured in the $\hat{\sigma}^z$ direction to obtain the imbalance $I(t)$. \textbf{c,} With the increase of system size, for 2D disordered many-body systems, the finite-size crossover point between the ETH and MBL regimes $W^{*}$ tends to a larger value.}\label{fig1}
\end{figure*}

Despite the MBL regime in finite-size 1D systems has been extensively investigated over the past decade, the status of MBL in two-dimensional (2D) systems remains far less well understood, primarily constrained by computational complexity. Numerically, evidence for an MBL regime in finite-size 2D systems has been found based on the analysis of l-bits phenomenology of MBL~\cite{PhysRevLett.126.180602} and the quantum-circuit description~\cite{Wahl:2019wv}. However, it has also been argued that the 2D MBL regime can be significantly delocalized by rare thermal regimes~\cite{PhysRevLett.121.140601,PhysRevB.95.155129,PhysRevLett.125.155701,PhysRevB.106.184209}. On the experimental front, signatures of slow relaxation dynamics have been observed in 2D ultracold atoms under strong disorder~\cite{doi:10.1126/science.aaf8834,PhysRevX.7.041047}, and a finite-size crossover regime between the ergodic and MBL regime has been probed from the Fock-space viewpoint in a 2D superconducting circuit~\cite{Yao:2023tc}. Nevertheless, due to the absence of a systematic finite-size analysis in the experimental studies, a central question remains unresolved: To what extent does the finite-size 2D MBL regime remain stable in the asymptotic limit of infinite system size?

\begin{figure*}[]
	\centering
	\includegraphics[width=1\linewidth]{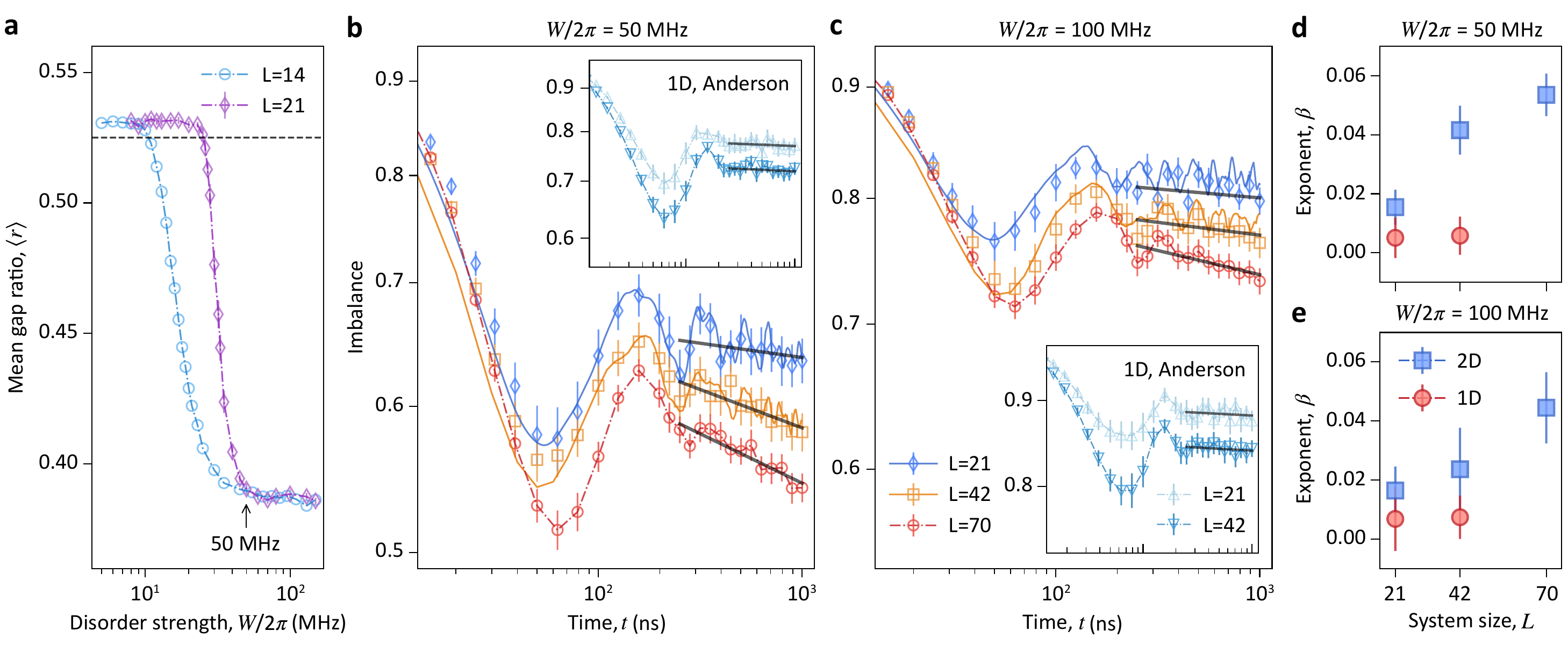}\\
	\caption{\textbf{Gap ratio and imbalance.} \textbf{a,} The mean gap ratio $\langle r \rangle$ as a function of disorder strength $W$ for both the $7\times 2$-qubit system ($L=14$), and the $7\times 3$-qubit system ($L=21$). The dashed horizontal line represents the threshold value $\langle r\rangle_{T} =0.525$. \textbf{b,} The dynamics of imbalance $I(t)$ for the disordered 2D systems with sizes $L=21$, $42$, $70$, and a disorder strength $W/2\pi = 50$ MHz. The inset shows the dynamics of $I(t)$ for disordered 1D systems  with sizes $L=21$, $42$, and a disorder strength $W/2\pi = 50$ MHz. The markers (connected by the dashed line) denote the experimental data with error bars, the solid curves show the numerical data, and the straight black lines represent the power-law fitting for the experimental data. Fig.~\textbf{c} is similar to Fig.~\textbf{b} but with a larger  disorder strength $W/2\pi = 100$ MHz. \textbf{d,} The relaxation exponent $\beta$ as a function of system sizes $L$ for both 2D and 1D systems with a disorder strength $W/2\pi = 50$ MHz. Fig.~\textbf{e} is similar to Fig.~\textbf{d} but with a larger  disorder strength $W/2\pi = 100$ MHz. Error bars on imbalance (Fig.~\textbf{b} and \textbf{c}) denote the standard error of the statistical mean, while those on relaxation exponent (Fig.~\textbf{d} and \textbf{e}) indicate the one standard error of the slopes obtained from the linear fits.}\label{fig2}
\end{figure*}

To address this open question, we perform an analog quantum simulation of the dynamics of the 2D $XY$ model with tunable chemical potential, using a large-scale superconducting circuit comprising 70 qubits and 117 couplers shown in Fig.~\ref{fig1}\textbf{a}. A schematic representation of the analog quantum simulation is displayed in Fig.~\ref{fig1}\textbf{b}. The high controllability of superconducting circuits allows the precise manipulation of the disorder, enabling us to tune the system between the ETH and MBL regimes. The scalability of the platform enables for the finite-size analysis of the ETH-MBL crossover. Recently, the robustness of the MBL regime has been experimentally studied by coupling the localized regime to a thermal inclusion, showing that the MBL regime can be delocalized by an artificially controlled rare ergodic inclusion~\cite{PhysRevX.9.041014,Leonard:2023vv}. Here, in contrast, we directly study the ETH-MBL crossover with different system sizes, observing a clear signature of the delocalization of finite-size MBL regime as the system size increases, consistent with the theoretical and numerical arguments~\cite{PhysRevB.99.134305,PhysRevLett.125.155701} rooted in the avalanche mechanism.

\emph{Experimental setup and protocol.}---We perform experiments by employing the programmable superconducting quantum simulator consisting of 70 transmon qubits with the 2D rectangular geometry (Fig.~\ref{fig1}\textbf{a}). All nearest-neighbor (NN) qubits are connected via tunable couplers~\cite{PhysRevApplied.10.054062}. We develop and implement several automation techniques, including automatic qubit bringing-up, allocation of idle points~\cite{cpl_40_6_060301} to ensure a stable qubit performance (average energy relaxation time of $\overline{T_1}$ = 52.3 $\mu$s) and the synchronization of control pulses~\cite{cpl_40_6_060301,PRXQuantum.6.010325} (see Supplementary Materials for details). Using these methods, we calibrate such large-scale quantum circuits into a qualified quantum simulator to realize a 2D $XY$ Hamiltonian
\begin{eqnarray}
\hat{H} =  \sum_{\langle i,j \rangle}\frac{J_{ij}}{2} (\hat{\sigma}_{i}^{x}\hat{\sigma}_{j}^{x} + \hat{\sigma}_{i}^{y}\hat{\sigma}_{j}^{y}) + \sum_{i} h_{i} \hat{\sigma}_{i}^{+}\hat{\sigma}_{i}^{-}, 
\label{hamiltonian}
\end{eqnarray}
where $J_{ij}$ represents the couplings between $i$-th and $j$-th qubits, $h_{i}$ denotes tunable on-site chemical potentials, and  $\hat{\sigma}_{i}^{x}$, $\hat{\sigma}_{i}^{y}$, $\hat{\sigma}_{i}^{\pm} = (\hat{\sigma}_{i}^{x} \pm i\hat{\sigma}_{i}^{y})/2$ are Pauli operators acting in the $i$-th qubit. In this experiment, we apply on-site potentials $h_{i}$ drawn uniformly from the interval $[-W,W]$, where $W$ is the disorder strength. We control the on-site potentials and coupling strength via well-corrected Z pulses on both qubits and couplers, compensating pulse distortions as described in Ref.~\cite{doi:10.1126/science.aaw1611, PhysRevApplied.23.024059}, thereby ensuring faithful evolution of the target Hamiltonian. To calibrate the exact bare qubit frequencies~\cite{Andersen:2025ut} (rather than coupling-shifted dressed frequencies), we perform the time-domain Rabi oscillations, along with the detailed qubit spectroscopy~\cite{PRXQuantum.6.010325,PhysRevLett.131.080401}. These methods allow us to precisely set on-site potentials and disorder strengths within a 100 MHz window (further details provided in the Supplementary Materials).

There are two main contributions of the couplings in our quantum simulator: nearest-neighbor qubits couplings $J_{ij}^\mathsmaller{\rm NN}$, realized via tunable couplers with a mean value of $\smash{\overline{J_{ij}^\mathsmaller{\rm NN}}}/2\pi \simeq 2.9$ MHz, and additional next-nearest-neighbor qubits couplings $J_{ij}^\mathsmaller{\rm NNN}$ with a mean value of $\smash{\overline{J_{ij}^\mathsmaller{\rm NNN}}}/2\pi \simeq 1.1$ MHz (see Fig.~\ref{fig1}\textbf{b}). All couplings are calibrated via two-qubit swap spectroscopy~\cite{Shi:2023un}. The calibration accuracy of both on-site potentials and coupling strengths is validated by the agreement between single-qubit populations measured under on-resonant dynamics and those obtained from exact numerical simulations across multiple block segments of the quantum processor (see Supplementary Materials for details).

To study the stability of the MBL regime with increasing system sizes, as shown in Fig.~\ref{fig1}\textbf{a}, we consider three different systems with sizes $L=21$ ($3 \times 7$ square-lattice qubit array), $L=42$ ($6 \times 7$ square-lattice qubit array), and $L=70$ with all available qubits in the device (marked by orange circles in Fig.~\ref{fig1}\textbf{a}). Each qubit is initially prepared in the vacuum state $|0\rangle$. Then, we select $n=L/2$ qubits (for $L=21$, we consider both $n=10$ and $11$) and implement $X$ gate on these qubits to excite them to the state $|1\rangle$, which results in a half-filling ``checkerboard" product state $|\psi_{0}\rangle$ (see Supplementary Material for the detailed structure). Subsequently, we implement the unitary dynamics $|\psi_{t}\rangle = \exp(-i \hat{H} t) |\psi_{0}\rangle$ with $\hat{H}$ being the Hamiltonian (\ref{hamiltonian}) by modulating both qubit and coupler frequencies. Finally, we measure the imbalance of the non-equilibrium state $|\psi_{t}\rangle$, defined as $I(t) = (\langle n_{1}(t)\rangle - \langle n_{0}(t)\rangle)/(\langle n_{1}(t)\rangle +\langle n_{0}(t)\rangle )$, where
\begin{eqnarray}
\langle n_{m}(t)\rangle = \frac{1}{L_{m}}\sum_{j\in{|m\rangle}} \langle\psi_{t} | \hat{n}_{j} |\psi_{t}\rangle 
\label{n}
\end{eqnarray}
for $m=0,1$, with $\hat{n}_{j} = \hat{\sigma}_{i}^{+}\hat{\sigma}_{i}^{-}$ being the particle-number operator, $j\in{|m\rangle}$ denoting the sites initialized by the state $| m \rangle$, and $L_{m}$ the number of sites with the initial state $| m \rangle$. The imbalance quantifies the loss of memory of initial states during relaxation dynamics, allowing us to probe whether the system thermalizes. For sufficiently strong disorder, a subdiffusive power-law decay of the imbalance, i.e., $I(t)\propto t^{-\beta}$ with $\beta < 0.5$, has been widely observed numerically~\cite{PhysRevLett.125.155701,PhysRevB.106.184209,PhysRevLett.114.160401,PhysRevB.93.060201,PhysRevB.96.075146, PhysRevB.105.224203} and experimentally~\cite{PhysRevX.7.041047,PhysRevLett.119.260401,Guo:2021wr,Shi:2024wm}.

\emph{Gap ratio analysis.}---We start by numerically studying the properties of ETH-MBL crossover in our system (\ref{hamiltonian}) by adopting the mean gap ratio $\langle r \rangle$~\cite{PhysRevB.75.155111}, as a conventional probe of the MBL regime. The mean gap ratio $\langle r \rangle$ is the average of $r_{n} = \min(\delta_{n},\delta_{n-1})/\max(\delta_{n},\delta_{n-1})$ with $\delta_{n} = E_{n+1} - E_{n}$ and $E_{n}$ being the ordered eigenvalues of the Hamiltonian (\ref{hamiltonian}). For the 21-qubit system with the dimension of Hilbert space $\mathcal{N} = 352716$, we compute central 2000 eigenvalues for 60 realizations by using the polynomially filtered exact diagonalization method~\cite{PhysRevLett.125.156601} (see Supplementray Materials for details). We also calculate the $\langle r \rangle$ for a smaller system with $L=14$ ($7\times 2$-qubit ladder) for reference by using exact diagonalization with 500 realizations.

The results of $\langle r \rangle$ are shown in Fig.~\ref{fig2}\textbf{a}. For ergodic systems, the energy level statistics follow the Gaussian orthogonal ensemble (GOE), and the mean gap ratio is $\langle r \rangle_{\text{GOE}} \simeq 0.53 $. Here, we choose a threshold $\langle r\rangle_{T} =0.525$ and estimate the boundary of ergodic regime, denoted by $W_{E}$, defined by the condition that for all disorder strengths  $W > W_{E}$, the gap ratio fulfills $\langle r \rangle <\langle r\rangle_{T}$. 
For small system of $L=7\times 2$ qubits, we observe that $W_{E}/2\pi \simeq 10$ MHz, while for larger system $L=7\times 3$, the ergodic regime extends already up to $W_{E}/2\pi \simeq 25$ MHz,
exhibiting a clear tendency towards ergodic behavior with increasing system size $L$.
Moreover, there is a crossing point between the curves of $\langle r\rangle$ as a function of $W$ for different system sizes near $W_{c}/2\pi \simeq 50$ MHz. At the early stages of MBL studies, the crossing point $W_{c}$ was regarded as an estimation of the transition point between the ergodic and MBL \emph{phases}~\cite{PhysRevB.82.174411,PhysRevX.7.021013,PhysRevB.91.081103,PhysRevLett.119.075702,PhysRevLett.126.100604,PhysRevB.107.115132}, while recent studies clarify that in 1D systems, the MBL \emph{phase} can occur at larger disorder strengths $W \gg W_{c}$~\cite{PhysRevB.105.174205, PhysRevLett.125.155701}. Consequently, based on the analysis of $\langle r \rangle$, one can see that in our superconducting circuit, the disorder strength $W_{c}/2\pi \simeq 50$ MHz can be regarded as an estimation of the lower bound of the MBL \emph{regime}.

\emph{Results of imbalance.}---We now study the dynamics of imbalance $I(t)$. In Fig.~\ref{fig2}\textbf{b} and~\ref{fig2}\textbf{c}, we plot the experimental data for the $I(t)$ with different system sizes and disorder strengths $W/2\pi = 50$ MHz and $100$ MHz, respectively. At the timescale $t\simeq 1000$ ns, we observe a slow decay of imbalance well approximated by a power law, i.e., $I(t)\propto t^{-\beta}$. Notably, for both $W/2\pi = 50$ MHz and $100$ MHz, the relaxation exponent $\beta$ increases with system size $L$. 
These results suggest that even at disorder strengths above the estimated crossing point $W_{c}/2\pi \simeq 50$ MHz (based on the $\langle r \rangle$ data in Fig.~\ref{fig2}\textbf{a}), the finite-size MBL regime becomes unstable and exhibits progressively stronger delocalization as $L$ increases.

For the dynamics of $I(t)$ with the system size $L=7\times 3$ and $7\times 6$, 
we also perform numerical simulations by using the Krylov subspace method and matrix-product-state based time-dependent variational principle (TDVP) algorithm (see Supplementary Materials for details). The computational cost of accurately simulating the dynamics of 2D disordered systems is substantial. In particular, TDVP simulations at a disorder strength of $W/2\pi = 100$ MHz require bond dimensions exceeding $\chi = 1000$ to achieve convergence (see Supplementary Materials). The experimental data are well consistent with the numerical results (see Fig.~\ref{fig2}\textbf{b} and~\ref{fig2}\textbf{c}), indicating that the experimental errors remain well controlled on the considered timescales. We also experimentally study the dynamics of $I(t)$ in disordered 1D qubit chains with size $L=21$ and $42$, realizing the 1D disordered $XY$ model (see the inset of Fig.~\ref{fig2}\textbf{b} and~\ref{fig2}\textbf{c}). This model exhibits Anderson localization and the relaxation exponent of imbalance is theoretically predicted to be $\beta = 0$~\cite{PhysRev.109.1492,PhysRevLett.125.155701} at any considered system size.

In Fig.~\ref{fig2}\textbf{d} and~\ref{fig2}\textbf{e}, we plot the relaxation exponent $\beta$ as a function of $L$ for $W/2\pi = 50$ MHz and $100$ MHz, respectively. We estimate the exponent $\beta$ by fitting the imbalance $I(t)\propto t^{-\beta}$ with a time interval $t\in [250,1000]$ ns. For 2D disordered systems, even at disorder stengths $W/2\pi\geq 50$ MHz, the exponent $\beta$ increases linearly with system size $L$, indicating that in this regime, although thermalization is extremely slow, it is not completely arrested, and many-body delocalization eventually occurs. In clear contrast to the 2D many-body case, there is no significant increase of the relaxation exponent $\beta$ with system size $L$ for the case of 1D Anderson localization.

\begin{figure}[t]
	\centering
	\includegraphics[width=1\linewidth]{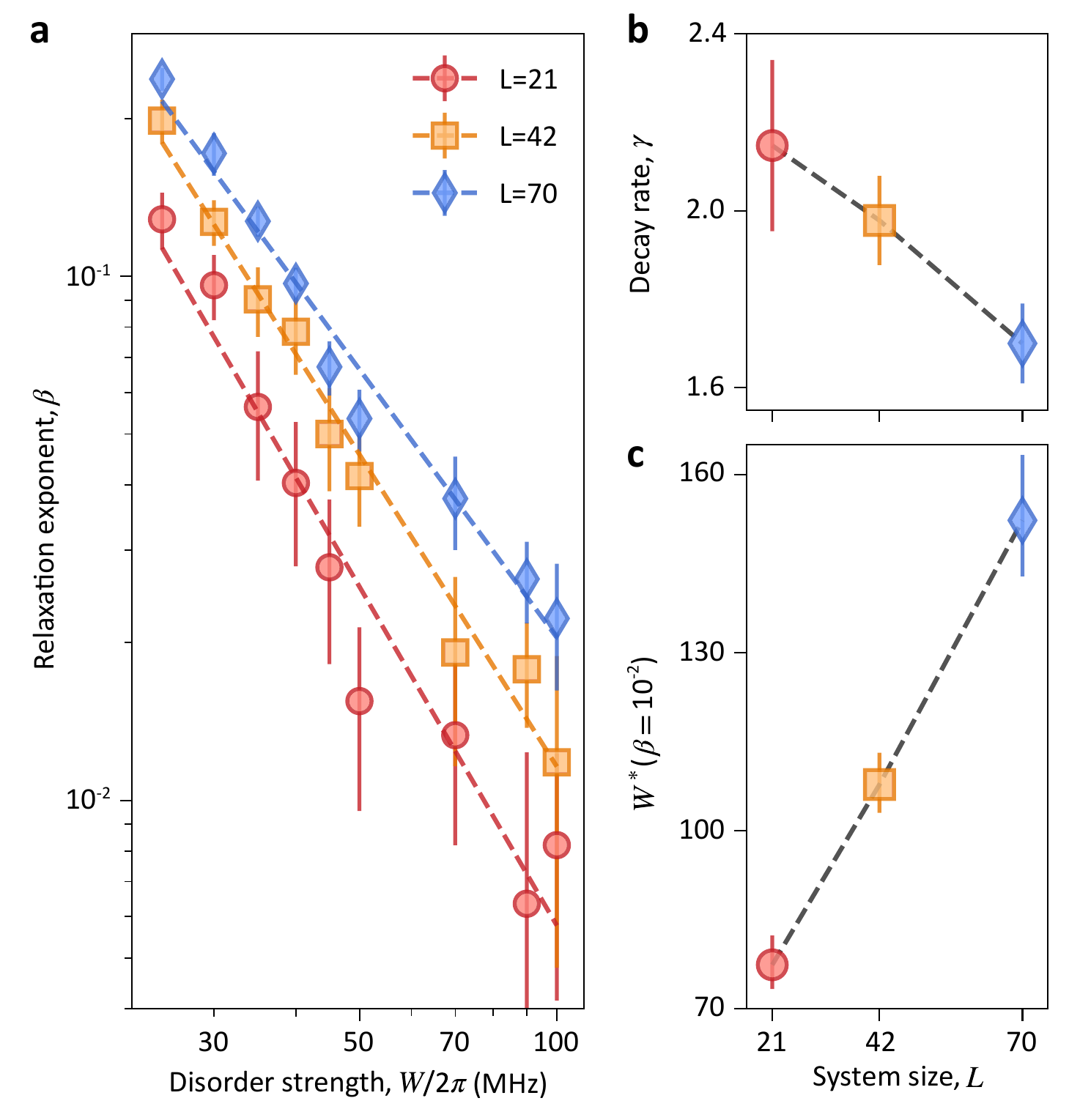}\\
	\caption{\textbf{Many-body delocalization.}  \textbf{a,} The relaxation exponent $\beta$ obtained by fitting the experimental data of imbalance as a function of disorder strength $W$ in 2D systems with different sizes $L$. The dashed lines represent the fittings of $\beta$ by considering the function $\beta = C W^{-\gamma}$ with two fitting parameters $C$ and $\gamma$, which satisfies $\beta\rightarrow 0$ when $W\rightarrow \infty$. \textbf{b,} The value of $\gamma$ as a function of $L$. Error bars on both the data of $\gamma$ denote the uncertainty of the fit. \textbf{c,} The estimated value of $W^{*}(\beta=10^{-2})$ based on the fittings in \textbf{a} as a function of $L$. Error bars on $\beta$ and $\gamma$ denote one standard error obtained from the fits. Error bars on $W^{*}(\beta=10^{-2})$ denote one standard deviation estimated from $N_{\text{rep}}=5000$ samples drawn from the joint Gaussian distribution of the fitted intercept and slope.}\label{fig3}
\end{figure}

\emph{Evidence of many-body delocalization.}---Next, we plot the exponent $\beta$ for experimental data as a function of disorder strength $W$ with different system sizes $L$ in Fig.~\ref{fig3}\textbf{a}. Although in 1D disorder Heisenberg model, it has been shown that $\beta$ decays \emph{exponentially} with $W$~\cite{Sierant_2025}, here we find that a \emph{power-law} decay fits better for the  $\beta$ as a function of $W$, i.e., $\beta \propto W^{-\gamma}$ (see Supplementary Materials for the comparison between the power-law and exponential fittings). 

In  Fig.~\ref{fig3}\textbf{b}, we display the decay rate $\gamma$ with different system sizes $L$. The significant decrease of $\gamma$ with system sizes indicates that the relaxation exponent $\beta$ becomes more robust against the increase of disorder strength $W$ for systems with larger sizes. Although the results of mean gap ratio $\langle r \rangle$ in Fig.~\ref{fig2}\textbf{a} show that $\langle r \rangle$ satisfies the value of Poisson distribution $\langle r \rangle_{\text{Poisson}}\simeq 0.39$ with $W/2\pi \gtrsim 50$ MHz, many-body delocalization in 2D system occurs up to a relatively large disorder strength of $W/2\pi = 100$ MHz ($W/J\simeq 34.5$).

To pinpoint the disorder strength at which the the relaxation dynamics essentially halts, we set a threshold $\beta = 10^{-2}$ to identify the disorder strength $W^{*}(\beta=10^{-2})$. The reason behind the choice of the threshold is that based on the experimental data of 1D Anderson localization, the measured relaxation exponents $\beta$ are on the order of $10^{-2}$ (see Fig.~\ref{fig2}\textbf{d} and~\ref{fig2}\textbf{e}). The value of $W^{*}(\beta=10^{-2})$ is extracted from the fitting curve in Fig.~\ref{fig3}\textbf{a}. In Fig.~\ref{fig3}\textbf{c}, we plot $W^{*}(\beta=10^{-2})$ as a function of system size $L$. We observe an approximate linear increase of $W^{*}(\beta=10^{-2})$ without bound up to the size $L=70$. These findings provide evidence that, in 2D systems, the delocalized regime challenges the finite-size MBL regime, and may ultimately prevail in the thermodynamic limit.

\emph{Discussion and outlook.}---We have experimentally explored the robustness of the MBL regime in 2D quantum many-body systems, focusing on time evolution of imbalance $I(t)$ as an indicator of ergodicity breaking. Our results provide evidence for a phenomenon dubbed many-body delocalization. At a fixed and relatively large disorder strength $W$, we observe non-ergodic behavior, manifested by the saturation of the imbalance at non-zero values --- signifying the onset of the MBL regime at a given system size $L$. However, this non-ergodic MBL regime is unstable: increasing the system size induces a persistent decay of $I(t)$, ultimately leading to the restoration of ergodicity for sufficiently large systems.
Specifically, for our 2D disordered superconducting quantum simulator, we observe slow thermalization up to a disorder strength of ${W/J}\approx{35}$. Extrapolations of our experimental data provide no evidence for the existence of an MBL \emph{phase} at any disorder strength in two-dimensional systems.

We note that, unlike one-dimensional systems, where tensor-network methods can probe the MBL regime out to relatively long timescales~\cite{PhysRevB.101.035148}, accurate numerical simulations of disordered two-dimensional dynamics, even at the strongest disorder, demand substantial classical computational resources. Consequently, the quantum simulations of disordered 2D systems presented here tackle a problem that is potentially intractable on classical hardware.

We have adopted the power-law decay of the imbalance to characterize the relaxation exponent in 2D systems. With the further improvement of the coherence in this platform, it is potential to observe a faster decay of imbalance than the power law~\cite{PhysRevLett.125.155701}, or a stretched exponential behavior proposed for the MBL regime which eventually thermalizes~\cite{PhysRevLett.131.106301}. 

Our work shows that superconducting circuits offer a scalable platform for studying intriguing phenomena emerging in the out-of-equilibrium dynamics of quantum many-body systems that remain widely unexplored beyond 1D, such as the Hilbert-space fragmentation~\cite{PRXQuantum.6.010325,PhysRevLett.133.196301,Adler:2024tg}, roughening dynamics~\cite{2024arXiv241210145K,2025arXiv250611187H}, and quantum coarsening~\cite{2024arXiv240115144S,Andersen:2025ut,Manovitz:2025uy}.

\begin{acknowledgments}

This work was supported by the National Natural Science Foundation of China (Grants Nos. 92265207, T2121001, 92365301, T2322030, 12122504, 12274142, 12475017), the Innovation Program for Quantum Science and Technology (Grant No. 2021ZD0301800), the Beijing Nova Program (No. 20220484121, 20240484652), the Natural Science Foundation of Guangdong Province (Grant No. 2024A1515010398), the China Postdoctoral Science Foundation (Grant No. GZB20240815). We acknowledge the support from Synergetic Extreme Condition User Facility (SECUF) in Huairou District, Beijing. P.S. acknowledges fellowship within the ``Generación D" initiative, Red.es, Ministerio para la Transformación Digital y de la Función Pública, for talent attraction (C005/24-ED CV1), funded by the European Union NextGenerationEU funds, through PRTR.

\end{acknowledgments}

%

\end{document}


\title{Supplementary Information for \\Many-body delocalization with a two-dimensional 70-qubit superconducting quantum simulator}

\author{Tian-Ming Li}
    \thanks{These authors contributed equally to this work.}
    \affiliation{Beijing National Laboratory for Condensed Matter Physics, Institute of Physics, Chinese Academy of Sciences, Beijing 100190, China}
    \affiliation{School of Physical Sciences, University of Chinese Academy of Sciences, Beijing 100049, China}

\author{Zheng-Hang Sun}
    \thanks{These authors contributed equally to this work.}
    \affiliation{Theoretical Physics \uppercase\expandafter{\romannumeral3}, Center for Electronic Correlations and Magnetism, Institute of Physics, University of Augsburg, D-86135 Augsburg, Germany}

\author{Yun-Hao Shi}
    \thanks{These authors contributed equally to this work.}
    \affiliation{Beijing National Laboratory for Condensed Matter Physics, Institute of Physics, Chinese Academy of Sciences, Beijing 100190, China}

\author{Zhen-Ting Bao}
    \thanks{These authors contributed equally to this work.}
    \affiliation{Beijing National Laboratory for Condensed Matter Physics, Institute of Physics, Chinese Academy of Sciences, Beijing 100190, China}
    \affiliation{School of Physical Sciences, University of Chinese Academy of Sciences, Beijing 100049, China}
    
\author{Yong-Yi Wang}
    \affiliation{Beijing National Laboratory for Condensed Matter Physics, Institute of Physics, Chinese Academy of Sciences, Beijing 100190, China}
    \affiliation{School of Physical Sciences, University of Chinese Academy of Sciences, Beijing 100049, China}

\author{Jia-Chi Zhang}
    \affiliation{Beijing National Laboratory for Condensed Matter Physics, Institute of Physics, Chinese Academy of Sciences, Beijing 100190, China}
    \affiliation{School of Physical Sciences, University of Chinese Academy of Sciences, Beijing 100049, China}

\author{Yu Liu}
    \affiliation{Beijing National Laboratory for Condensed Matter Physics, Institute of Physics, Chinese Academy of Sciences, Beijing 100190, China}
    \affiliation{School of Physical Sciences, University of Chinese Academy of Sciences, Beijing 100049, China}
    
\author{Cheng-Lin Deng}
    \affiliation{Beijing Key Laboratory of Fault-Tolerant Quantum Computing, Beijing Academy of Quantum Information Sciences, Beijing 100193, China}

\author{Yi-Han Yu}
    \affiliation{Beijing National Laboratory for Condensed Matter Physics, Institute of Physics, Chinese Academy of Sciences, Beijing 100190, China}
    \affiliation{School of Physical Sciences, University of Chinese Academy of Sciences, Beijing 100049, China}

\author{Zheng-He Liu}
    \affiliation{Beijing National Laboratory for Condensed Matter Physics, Institute of Physics, Chinese Academy of Sciences, Beijing 100190, China}
    \affiliation{School of Physical Sciences, University of Chinese Academy of Sciences, Beijing 100049, China}

\author{Chi-Tong Chen}
    \affiliation{Quantum Science Center of Guangdong-Hong Kong-Macao Greater Bay Area, Shenzhen, Guangdong 518045, China}

\author{Li Li}
    \affiliation{Beijing National Laboratory for Condensed Matter Physics, Institute of Physics, Chinese Academy of Sciences, Beijing 100190, China}
    \affiliation{School of Physical Sciences, University of Chinese Academy of Sciences, Beijing 100049, China}

\author{Hao Li}
    \affiliation{Beijing Key Laboratory of Fault-Tolerant Quantum Computing, Beijing Academy of Quantum Information Sciences, Beijing 100193, China}

\author{Hao-Tian Liu}
    \affiliation{Beijing National Laboratory for Condensed Matter Physics, Institute of Physics, Chinese Academy of Sciences, Beijing 100190, China}
    \affiliation{School of Physical Sciences, University of Chinese Academy of Sciences, Beijing 100049, China}
    \affiliation{Beijing Key Laboratory of Fault-Tolerant Quantum Computing, Beijing Academy of Quantum Information Sciences, Beijing 100193, China}

\author{Si-Yun Zhou}
    \affiliation{Beijing National Laboratory for Condensed Matter Physics, Institute of Physics, Chinese Academy of Sciences, Beijing 100190, China}
    \affiliation{School of Physical Sciences, University of Chinese Academy of Sciences, Beijing 100049, China}
    
\author{Zhen-Yu Peng}
    \affiliation{Beijing National Laboratory for Condensed Matter Physics, Institute of Physics, Chinese Academy of Sciences, Beijing 100190, China}
    \affiliation{School of Physical Sciences, University of Chinese Academy of Sciences, Beijing 100049, China}

\author{Yan-Jun Liu}
    \affiliation{Beijing National Laboratory for Condensed Matter Physics, Institute of Physics, Chinese Academy of Sciences, Beijing 100190, China}
    \affiliation{School of Physical Sciences, University of Chinese Academy of Sciences, Beijing 100049, China}

\author{Ziting Wang}
    \affiliation{Beijing Key Laboratory of Fault-Tolerant Quantum Computing, Beijing Academy of Quantum Information Sciences, Beijing 100193, China}

\author{Yueshan Xu}
    \affiliation{Beijing Key Laboratory of Fault-Tolerant Quantum Computing, Beijing Academy of Quantum Information Sciences, Beijing 100193, China}

\author{Kui Zhao}
    \affiliation{Beijing Key Laboratory of Fault-Tolerant Quantum Computing, Beijing Academy of Quantum Information Sciences, Beijing 100193, China}

\author{Yang He}
    \affiliation{Beijing National Laboratory for Condensed Matter Physics, Institute of Physics, Chinese Academy of Sciences, Beijing 100190, China}
    \affiliation{School of Physical Sciences, University of Chinese Academy of Sciences, Beijing 100049, China}

\author{Da'er Feng}
    \affiliation{Beijing National Laboratory for Condensed Matter Physics, Institute of Physics, Chinese Academy of Sciences, Beijing 100190, China}
    \affiliation{School of Physical Sciences, University of Chinese Academy of Sciences, Beijing 100049, China}

\author{Jia-Cheng Song}
    \affiliation{Beijing National Laboratory for Condensed Matter Physics, Institute of Physics, Chinese Academy of Sciences, Beijing 100190, China}
    \affiliation{School of Physical Sciences, University of Chinese Academy of Sciences, Beijing 100049, China}

\author{Cai-Ping Fang}
    \affiliation{Beijing National Laboratory for Condensed Matter Physics, Institute of Physics, Chinese Academy of Sciences, Beijing 100190, China}
    \affiliation{School of Physical Sciences, University of Chinese Academy of Sciences, Beijing 100049, China}
    \affiliation{Beijing Key Laboratory of Fault-Tolerant Quantum Computing, Beijing Academy of Quantum Information Sciences, Beijing 100193, China}

\author{Junrui Deng}
    \affiliation{Beijing National Laboratory for Condensed Matter Physics, Institute of Physics, Chinese Academy of Sciences, Beijing 100190, China}
    \affiliation{School of Physical Sciences, University of Chinese Academy of Sciences, Beijing 100049, China}

\author{Mingyu Xu}
    \affiliation{Beijing National Laboratory for Condensed Matter Physics, Institute of Physics, Chinese Academy of Sciences, Beijing 100190, China}
    \affiliation{School of Physical Sciences, University of Chinese Academy of Sciences, Beijing 100049, China}

\author{Yu-Tao Chen}
    \affiliation{Beijing National Laboratory for Condensed Matter Physics, Institute of Physics, Chinese Academy of Sciences, Beijing 100190, China}
    \affiliation{School of Physical Sciences, University of Chinese Academy of Sciences, Beijing 100049, China}

\author{Bozhen Zhou}
    \affiliation{Institute of Theoretical Physics, Chinese Academy of Sciences, Beijing 100190, China}

\author{Gui-Han Liang}
    \affiliation{Beijing National Laboratory for Condensed Matter Physics, Institute of Physics, Chinese Academy of Sciences, Beijing 100190, China}
    \affiliation{School of Physical Sciences, University of Chinese Academy of Sciences, Beijing 100049, China}

\author{Zhongcheng Xiang}
    \affiliation{Beijing National Laboratory for Condensed Matter Physics, Institute of Physics, Chinese Academy of Sciences, Beijing 100190, China}
    \affiliation{School of Physical Sciences, University of Chinese Academy of Sciences, Beijing 100049, China}

\author{Guangming Xue}
    \affiliation{Beijing Key Laboratory of Fault-Tolerant Quantum Computing, Beijing Academy of Quantum Information Sciences, Beijing 100193, China}

\author{Dongning Zheng}
    \affiliation{Beijing National Laboratory for Condensed Matter Physics, Institute of Physics, Chinese Academy of Sciences, Beijing 100190, China}
    \affiliation{School of Physical Sciences, University of Chinese Academy of Sciences, Beijing 100049, China}

\author{Kaixuan Huang}
    \affiliation{Beijing Key Laboratory of Fault-Tolerant Quantum Computing, Beijing Academy of Quantum Information Sciences, Beijing 100193, China}

\author{Zheng-An Wang}
    \affiliation{Beijing Key Laboratory of Fault-Tolerant Quantum Computing, Beijing Academy of Quantum Information Sciences, Beijing 100193, China}    
    
\author{Haifeng Yu}
    \email{hfyu@baqis.ac.cn}
    \affiliation{Beijing Key Laboratory of Fault-Tolerant Quantum Computing, Beijing Academy of Quantum Information Sciences, Beijing 100193, China}

\author{Piotr Sierant}
    \affiliation{Barcelona Supercomputing Center, Barcelona 08034, Spain}
    
\author{Kai Xu}
    \email{kaixu@iphy.ac.cn}
    \affiliation{Beijing National Laboratory for Condensed Matter Physics, Institute of Physics, Chinese Academy of Sciences, Beijing 100190, China}
    \affiliation{School of Physical Sciences, University of Chinese Academy of Sciences, Beijing 100049, China}
    \affiliation{Beijing Key Laboratory of Fault-Tolerant Quantum Computing, Beijing Academy of Quantum Information Sciences, Beijing 100193, China}
    \affiliation{Hefei National Laboratory, Hefei 230088, China}
    \affiliation{Songshan Lake Materials Laboratory, Dongguan, Guangdong 523808, China}

\author{Heng Fan}
    \email{hfan@iphy.ac.cn}
    \affiliation{Beijing National Laboratory for Condensed Matter Physics, Institute of Physics, Chinese Academy of Sciences, Beijing 100190, China}
    \affiliation{School of Physical Sciences, University of Chinese Academy of Sciences, Beijing 100049, China}
    \affiliation{Beijing Key Laboratory of Fault-Tolerant Quantum Computing, Beijing Academy of Quantum Information Sciences, Beijing 100193, China}
    \affiliation{Hefei National Laboratory, Hefei 230088, China}
    \affiliation{Songshan Lake Materials Laboratory, Dongguan, Guangdong 523808, China}

\maketitle
\beginsupplement
\tableofcontents
\newpage

\section{\label{sec:model}Experimental setup and device performance}

\subsection{Experimental setup}

\begin{figure}[ht!]
	\centering
	\includegraphics[width=1.00\linewidth]{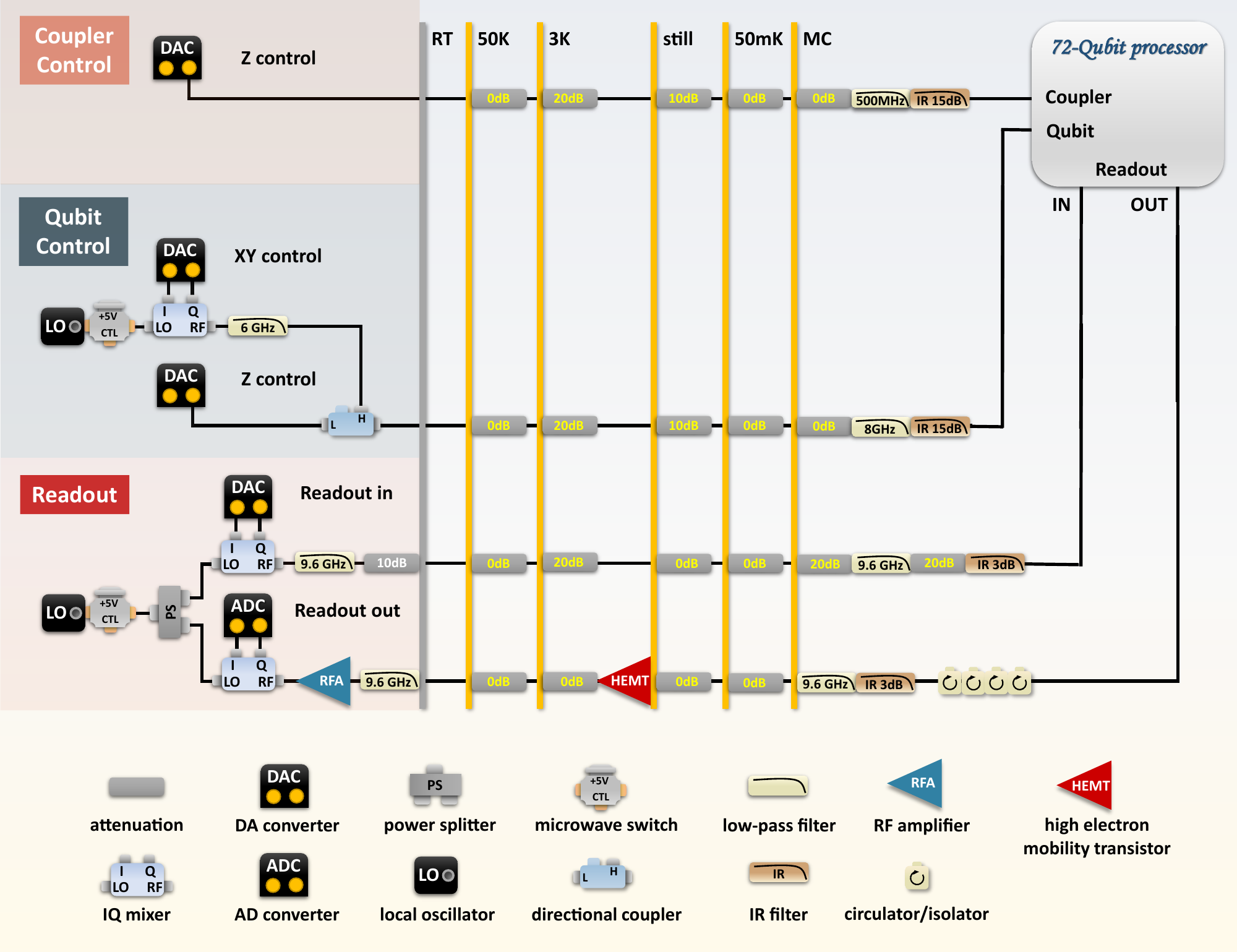}
	\caption{\textbf{Schematic diagram of electronic devices and  wiring information.}}
	\label{fig:S1A1}
\end{figure}

A cryogenic environment of $\sim$10 mK was realized and maintained using a Bluefors XLD-1000 dilution refrigerator to minimize thermal excitation for controlling superconducting circuits. As shown in Fig.\ref{fig:S1A1}, carefully selected attenuators and filters are arranged along the signal lines between room temperature (RT, $\sim$300 K) and the extremely low base temperature, effectively suppressing thermal noise and thereby protecting the quantum circuits (qubits and couplers).

High-frequency XY pulses ($\sim$ 4~GHz) and readout pulses ($\sim$ 7~GHz) are generated using in-phase/quadrature (IQ) modulation~\cite{Sank2014}. This technique combines the dual-channel arbitrary waveforms (frequencies below 500 MHz) generated by digital-to-analog converter (DAC) with microwave source (fixed at 3.350~GHz, 4.350~GHz, 6.675~GHz, 7.425~GHz in our experiment, respectively) produced by a local oscillator (LO). The intrinsic leakage from the LO signal is suppressed to approximately -90 dBm through zero calibration. The Z control for both qubits and couplers were generated by the similar DAC, which operating at 2 Gs/s with 16-bit vertical resolution. For couplers, the Z control signal is directly sent into the processor through coaxial cables and microwave components. In contrast, for qubits, the Z control signal is combined with the excitation pulse (XY signal) through a directional coupler prior to transmission. The joint readout signals are sent through the transmission line, sequentially amplified by a high electron mobility transistor (HEMT) and a room temperature RF amplifier (RFA), and finally demodulated by a dual-channel analog-digital converter (ADC) operating at 1 Gs/s with 14-bit vertical resolution. All electronic devices are synchronized via a clock-and-trigger generator paired with distributors. The generator is fed by a 10 MHz reference originated from the rubidium atomic clock, while the distributors up-convert this 10 MHz signal to 250 MHz and then disseminate it throughout the system. These distributors also route trigger signals to all DACs and ADCs.

\subsection{Device architecture}

\begin{figure}[ht!]
	\centering
	\includegraphics[width=1.00\linewidth]{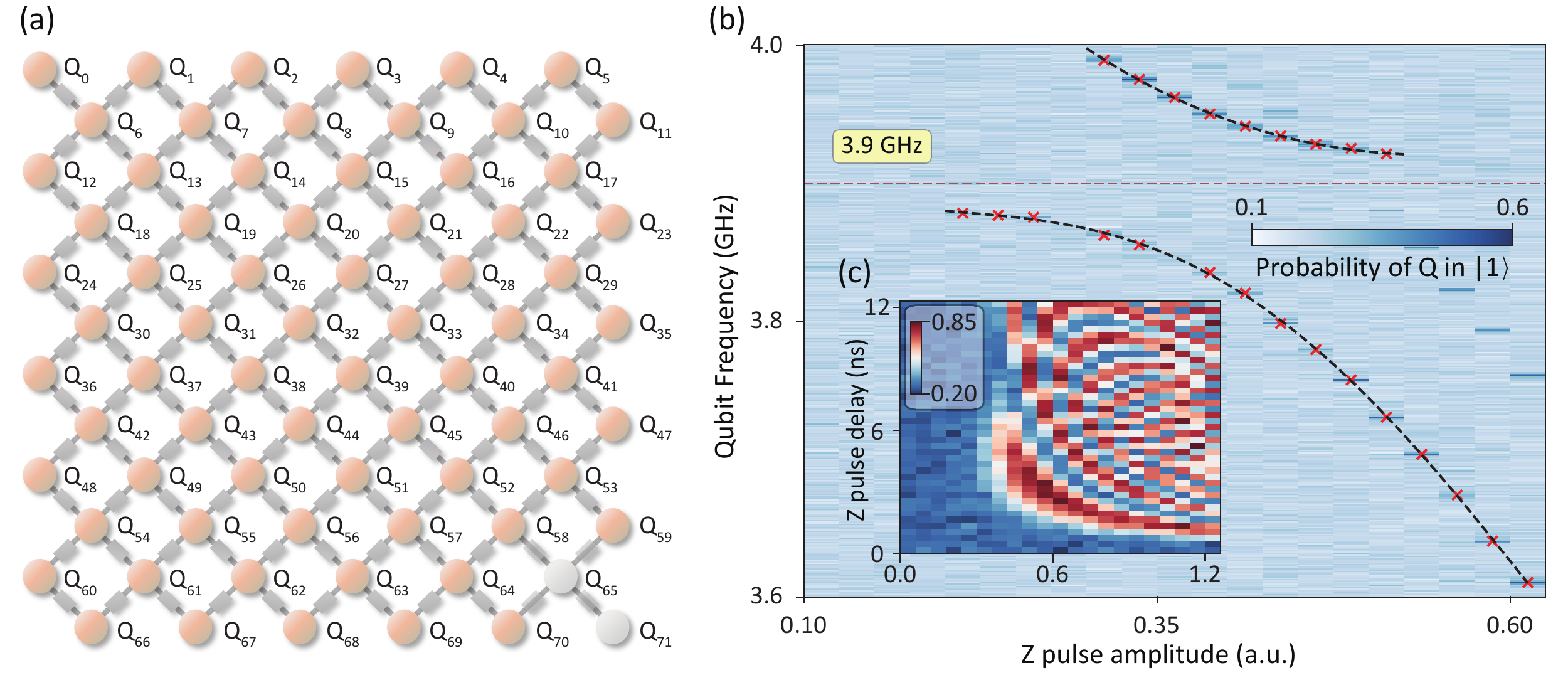}
	\caption{\textbf{Schematic of chip layout and characterizations of TLS strongly coupled to qubit $Q_{65}$.} (a) Layout of the quantum chip comprising a 2D array of 72 superconducting qubits and 121 couplers. Circles represent qubits, while rectangles represent couplers. Qubit $Q_{71}$ and 4 couplers (colored lighter gray) connected to $Q_{65}$ were not used in the experiments. Qubit labels follow a numbering sequence based on their physical locations. (b) Qubit spectroscopy of $Q_{65}$, showing a pronounced anti-crossing at approximately 3.9~GHz, indicative of strong coupling to a two-level system (TLS). The black dashed lines are fits to the upper and lower hybridized branches, while the red dashed line marks the TLS transition frequency. (c) Spectrally resolved $T_1$ of $Q_{65}$ in $\ket{0}$ after excitation, shown as the function of Z pulse amplitude. The persistent oscillations indicate coherent exchange between the qubit and the TLS.}
	\label{fig:S1B1}
\end{figure}

We conduct the experiments on a 72-qubit superconducting quantum processor arranged in 2D rectangular array (shown in Fig.\ref{fig:S1B1}(a)) similar to Sycamore~\cite{Arute2019}, including 12 readout lines, each coupled to 6 qubits. Each qubit is tunably coupled to its four nearest-neighbors (NN) typically, resulting in 121 tunable couplers used to flexibly control the effective coupling strength between qubits pairs.

In our experiments, only 70 qubits and 117 couplers on the processor are utilized. Qubit $Q_{65}$ is discarded due to its strong coupling with an external two-level system (TLS), which also results in the exclusion of $Q_{71}$, as it becomes effectively isolated. As shown in Fig.\ref{fig:S1B1}(b), the spectroscopy of $Q_{65}$ reveals a pronounced avoided crossing (anti-crossing), indicating the presence of a TLS at approximately 3.9~GHz with a coupling strength of about 50 MHz to $Q_{65}$. Fig.\ref{fig:S1B1}(c) further demonstrates the unwanted coherent exchange between $Q_{65}$ and the TLS, which adversely affects the experiment. Four couplers which connect to either $Q_{65}$ or $Q_{71}$ are therefore excluded from consideration.


\subsection{Device performance}

\begin{figure}[ht!]
	\centering
	\includegraphics[width=1.00\linewidth]{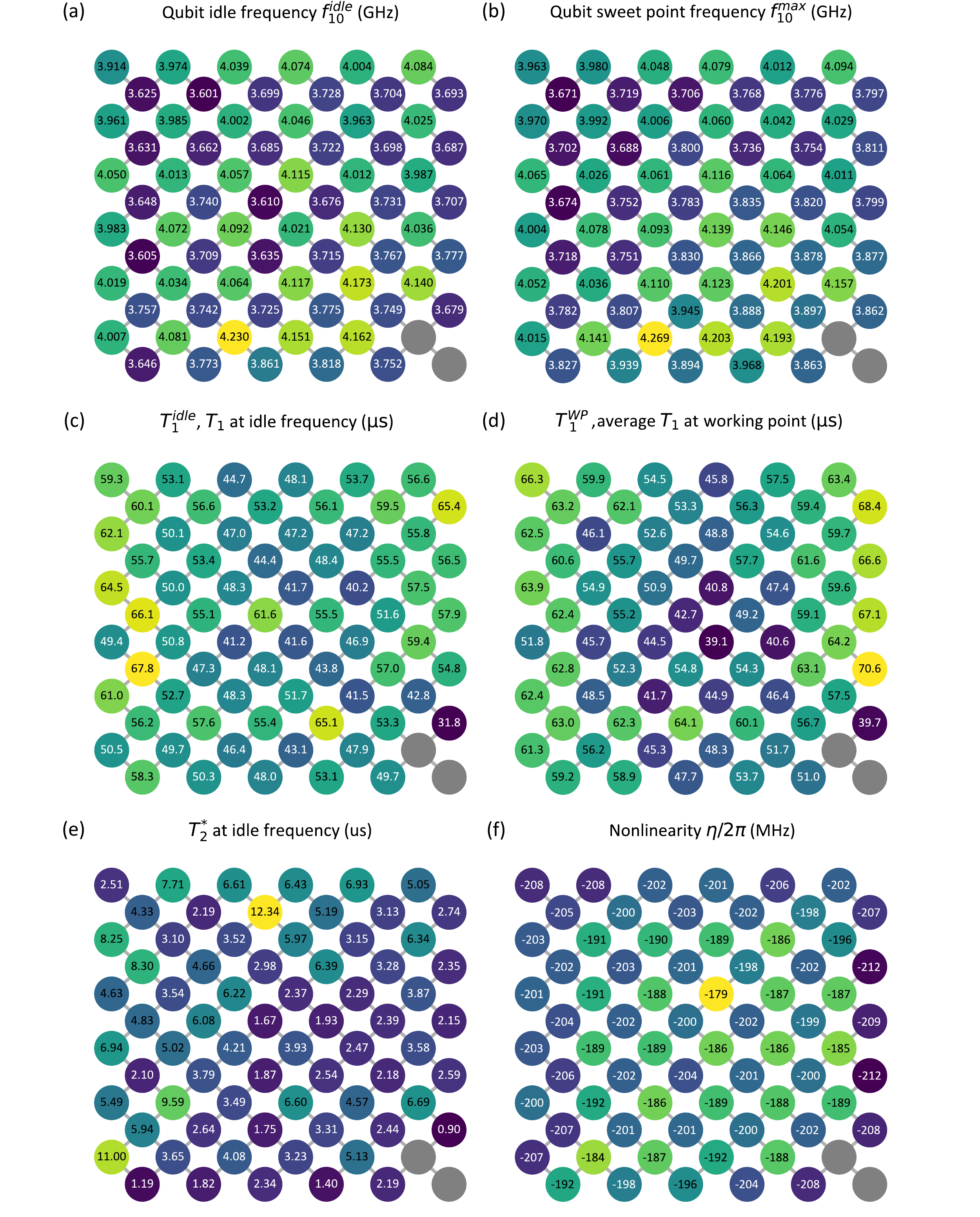}
	\caption{\textbf{Distribution of single-qubit performance parameters across the processor.} (a) Qubit $\ket{0}$ to $\ket{1}$ transition frequency at idle point, $f_{\rm 10}^{\rm{idle}}$. (b) Sweet point (maximum transition) frequency, $f_{\rm 10}^{\rm max}$. (c) Energy relaxation time at idle point, $T_1^{\rm idle}$ and (d) average energy relaxation time near resonant working point $T_1^{\rm WP}$. (e) Ramsey dephasing time, $T_{2}^*$ at idle point. (f) Qubit anharmonicity, $\eta/2\pi$. In all panels, black entries indicate values above the mean, while white entries denote values below the mean.}
	\label{fig:S1C1}
\end{figure}

\begin{figure}[ht!]
	\centering
	\includegraphics[width=1.00\linewidth]{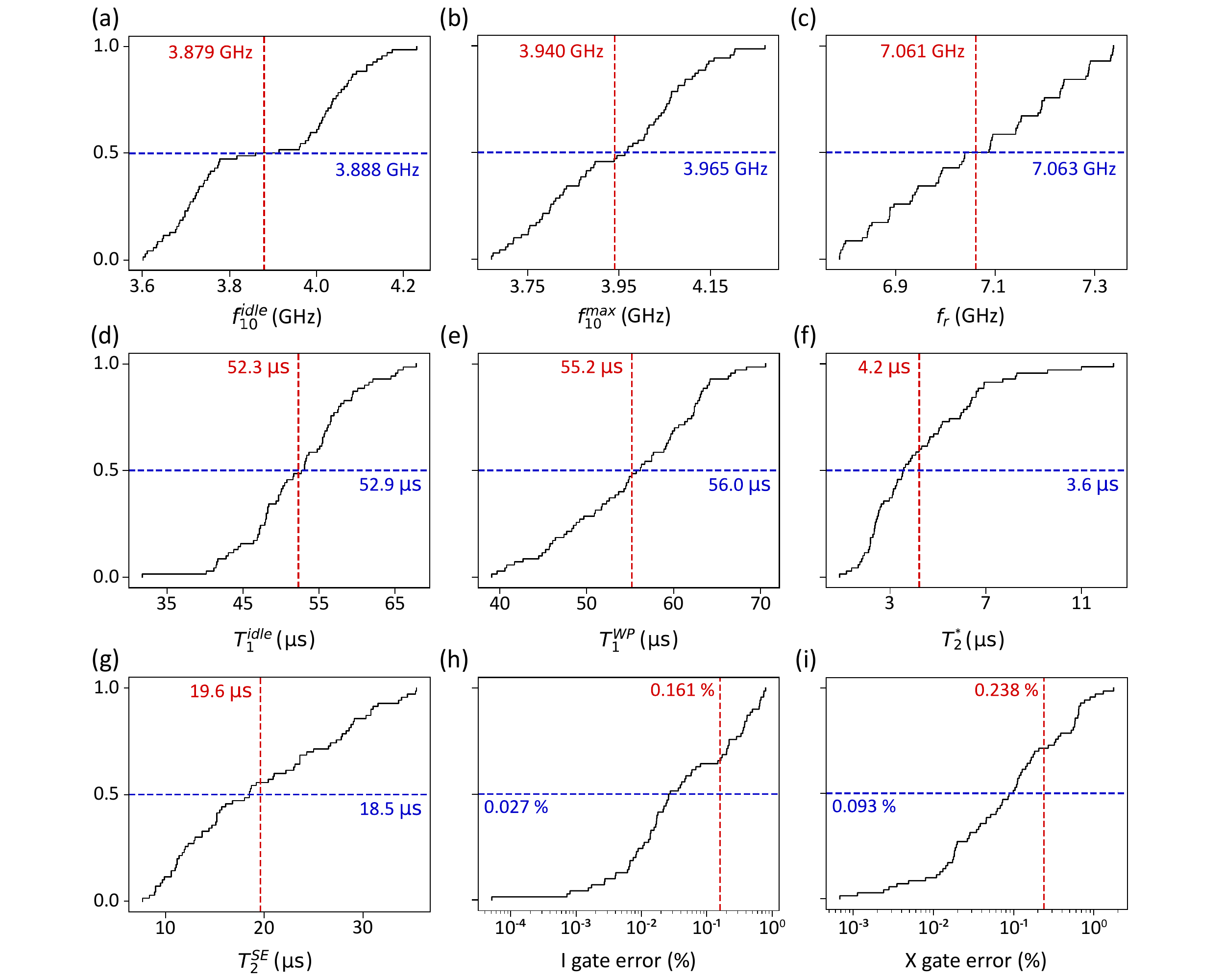}
	\caption{\textbf{Integrated histrogram of chip performance.} (a) Qubit $\ket{0}$ to $\ket{1}$ transition frequency at idle point, $f_{\rm 10}^{\rm{idle}}$. (b) Sweet point (maximum transition) frequency, $f_{\rm 10}^{\rm max}$. (c) Readout frequency, $f_r$. (d) Energy relaxation time at idle point, $T_1^{\rm idle}$, and (e) average energy relaxation time near resonant working point, $T_1^{\rm WP}$. (f) Ramsey dephasing time, $T_{2}^{\ast}$, and (g) spin-echo dephasing time, $T_{2}^{\rm SE}$, at idle point. Single-qubit (h) $I$ gate error rate and (i) $X$ gate error rate. Red vertical lines indicate mean values, while blue horizontal lines denote median values.}
	\label{fig:S1C2}
\end{figure}

In summarize, Fig.\ref{fig:S1C1} and Fig.\ref{fig:S1C2} list the basic performance of the qubits, including:

\begin{itemize}
    \item Qubit frequency, including $\ket{0}$ to $\ket{1}$ transition frequency at the idle point $f_{\rm 10}^{\rm{idle}}=\omega^{\rm{idle}}_{10}/2\pi$.
    \item Qubit sweet point (maximum) transition frequency $f_{\rm 10}^{\rm sw}=f_{\rm 10}^{\rm max}=\omega^{\rm{max}}_{10}/2\pi$.
    \item Qubit anharmonicity $\eta/2\pi=f_{21} - f_{10} < 0$, where $f_{21}$ is $\ket{1}$ to $\ket{2}$ transition frequency.
    \item Energy relaxation time at idle point $T_1^{\rm{idle}}$.
    \item Average energy relaxation time in the vicinity of resonant working point $T_1^{\rm{WP}}$ (range of $3.520\pm0.100 $~GHz).
    \item Dephasing time, including Ramsey dephasing $T_2^*$ and spin-echo dephasing $T_2^{\rm{SE}}$ at the idle point.
    \item Readout resonator frequency $f_r$.
    \item Single-qubit gate error rates for $I$ and $X$ gates measured using quantum state tomography (QST).
\end{itemize}

The detailed average performances of these parameters are displayed using vertical and horizontal lines shown in Fig.\ref{fig:S1C2}. We emphasize that the average energy relaxation time of both idle points and resonant working points are above 50 $\mu$s. 

\subsection{\label{sec:1D}Effective coupling strength}

\begin{figure}[ht!]
	\centering
	\includegraphics[width=1.00\linewidth]{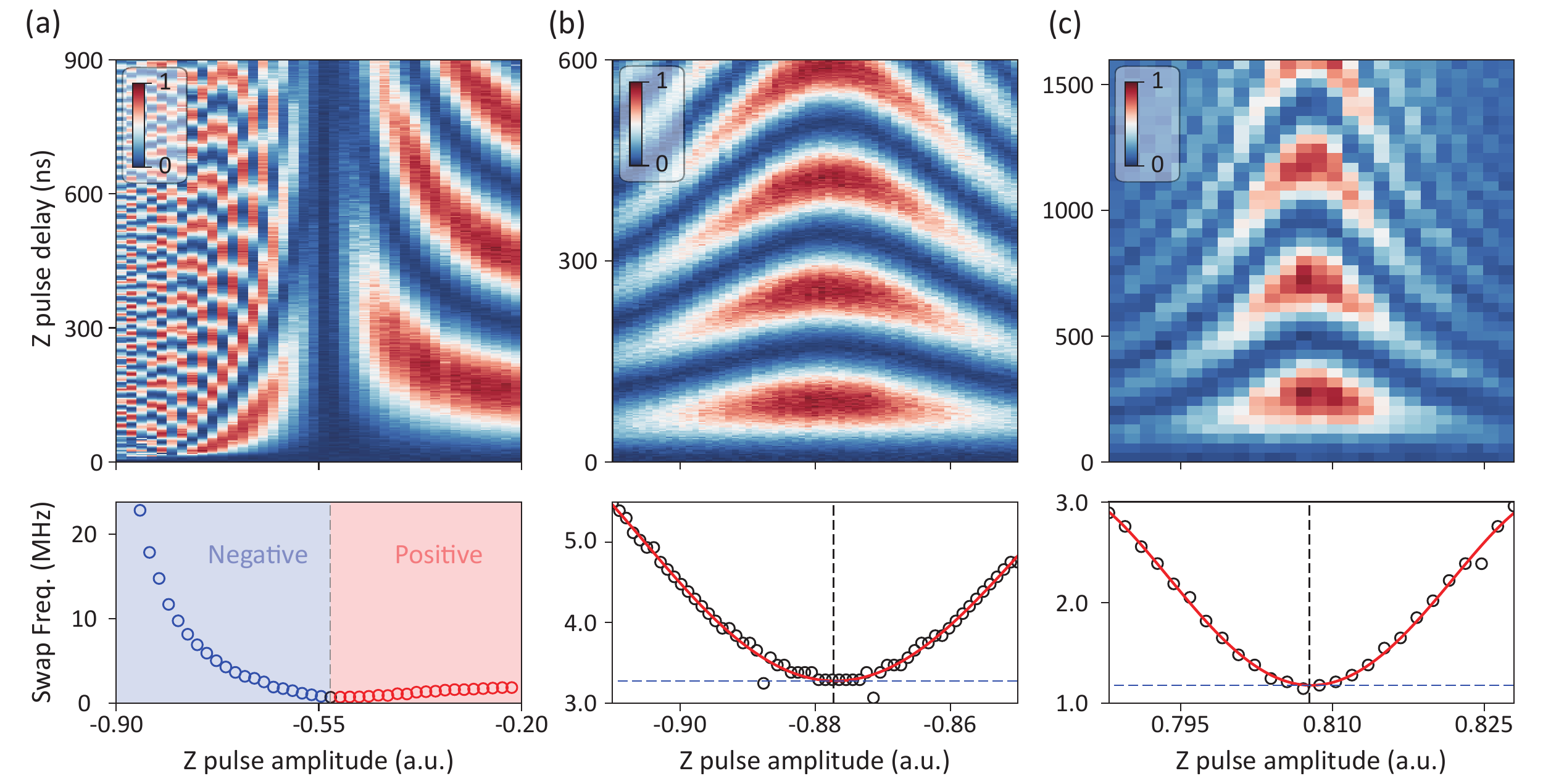}
	\caption{\textbf{Modulation and accurate measurement of effective coupling strength using two-qubit swap spectroscopy.} The top row shows the time-dependent probability of the qubit pair occupying $\ket{01}$ after initialization in $\ket{10}$, while the bottom row displays the corresponding swap frequencies obtained via Fourier transform. (a) Two NN qubits are both biased at 3.520~GHz, while the tunable coupler bias is varied via scanned Z pulse amplitude. In lower panel, the shaded red and blue regions denote positive and negative effective coupling strengths, respectively. Precise measurement of effective coupling strength for (b) an NN qubit pair and (c) a horizontally adjacent NNN qubit pair. One qubit remains fixed at 3.520~GHz, the other one is swept through resonance, while the coupler bias holds constant for fixing coupling strength. The effective coupling strength is extracted by fitting the minimum swap frequency observed in the Fourier transformed spectra.}
	\label{fig:S1D1}
\end{figure}

To realize tunable interactions between qubits, we adopt the Qubit–Coupler–Qubit (QCQ) architecture~\cite{Yan2018}. In this architecture, two NN qubits with frequencies $\omega_1$ and $\omega_2$, coupled directly with strength $g_{12}$, are principally coupled to an intermediate tunable coupler with frequency $\omega_c$, with coupling strengths $g_{1c}$ and $g_{2c}$, respectively. The effective coupling strength is then given by~\cite{Shi2023}
\begin{equation}
    g_{\rm{eff}} = g_{12} + \frac{g_{1c}g_{2c}}{2}\left(\frac{1}{\omega_1 - \omega_c} + \frac{1}{\omega_2 - \omega_c}\right),
\end{equation}
which can be adjusted by applying a suitable Z pulse amplitude to the coupler to modulate its frequency $\omega_c$. Fig.\ref{fig:S1D1}(a) shows the typical experimental dependence of the effective coupling strength $g_{\rm{eff}}$ as a function of the coupler Z pulse amplitude.

\begin{figure}[ht!]
	\centering
	\includegraphics[width=1.00\linewidth]{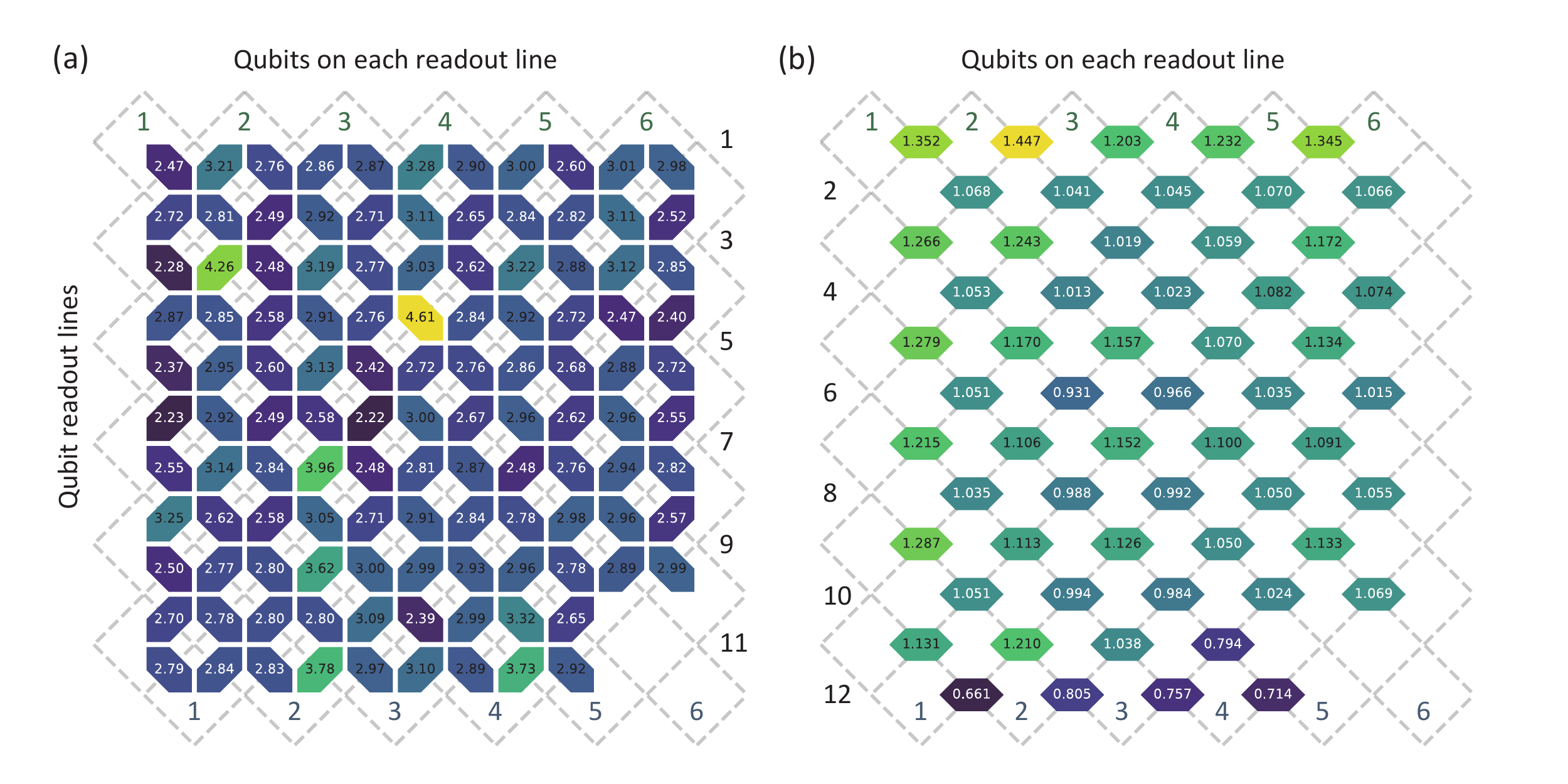}
	\caption{\textbf{Effective coupling strengths.} Effective coupling strengths (in MHz) are shown for (a) NN qubits and (b) horizontally adjacent qubits. In each panel, black entries denote values above the mean coupling strength, while white entries denote values below the mean.}
	\label{fig:S1D2}
\end{figure}

In our experiments, we characterize the effective coupling strengths between all NN qubit pairs [Fig.\ref{fig:S1D2}(a), typical data in Fig.\ref{fig:S1D1}(b)], as well as next-nearest-neighbour (NNN) qubit pairs in horizontal direction [Fig.\ref{fig:S1D2}(b), typical data in Fig.\ref{fig:S1D1}(c)], where the resonance point is set at 3.520~GHz. Due to the asymmetric floating layout of the coupler~\cite{sete2021}, horizontal qubit pairs are placed physically closer than vertical pairs. As a result, the coupling strengths are significantly different: horizontal pairs exhibit a typical strength around 1 MHz, while vertical pairs are weakly coupled by strength around 0.1 MHz and can be considered effectively decoupled.

\section{\label{sec:exp}Device control and calibration}

Controlling superconducting quantum processor with 70 qubits and 117 couplers is a highly complex task, posing some significant challenges in achieving a well-calibrated operational state across all qubits and couplers. This is crucial for accurately simulating dynamics related to system imbalances. Therefore, an automated and standardized calibration procedure is essential to ensure high-precision and stable control throughout the entire chip. In particular, when simulating the targeted many-body Hamiltonian, control parameters, such as flux bias used to tune qubits from their idle points to specific resonant working points, are required to be carefully optimized and validated. 

In this section, we outline the experimental details into two main parts: (i) \textit{automation technique}, which facilitates the initialization of our device on both the single-qubit and multi-qubit levels, and (ii) \textit{precision calibration}, which enables the accurate simulation for quantum many-body systems.

\subsection{Automation technique for scalable processor}
\subsubsection{\label{sec:2A1}Automatic qubit bringing-up}

\begin{figure}[ht!]
	\centering
	\includegraphics[width=1.00\linewidth]{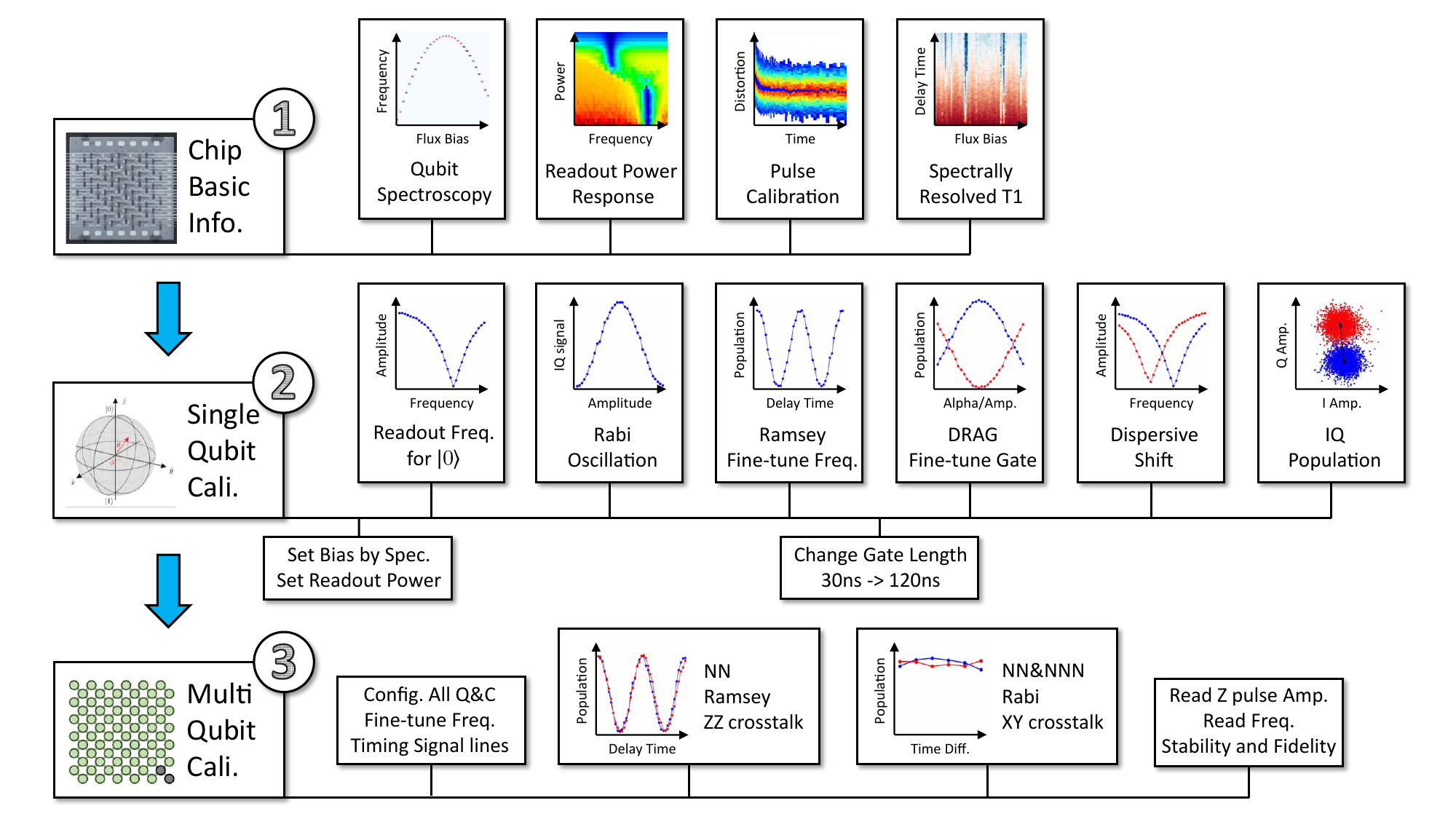}
	\caption{\textbf{Flowchart and representative data for automation techniques.} Part 1: Basic characterization of the chip, including (from left to right) qubit spectroscopy, readout signal amplitude as a function of drive frequency and power, flux pulse distortion and correction, and qubit spectrally resolved $T_1$. Part 2: Automated single-qubit bring-up. Sequential calibration steps are shown, along with corresponding representative data (see \ref{sec:2A1} for details). Part 3: Multi-qubit calibration. Workflow diagram illustrating how automated procedures establish a stable idle configuration with high-fidelity, stable readout and suppressed crosstalk (details in \ref{sec:3A2}).}
	\label{fig:S2A1}
\end{figure}

Following a round of initial calibration across the entire processor, we obtain some basic information, typical data of which is displayed in Fig.\ref{fig:S2A1}:
\begin{itemize}
    \item Qubit spectroscopy, which establishes the relationship between flux bias and transition frequency.
    \item The response of the readout signal amplitude as a function of drive frequency and amplitude~\cite{Bishop2010, Boissonneault2010}, which determines the appropriate readout power for each qubit.
    \item Flux signal correction~\cite{Yan2019, Li2025}, achieved by predistorting Z pulse using the step response function measured at a calibration working point approximately 0.8~GHz below the sweet point, thereby providing enhanced resolution due to the increased sensitivity of the flux bias.
    \item Qubit spectrally resolved $T_1$ (i.e., spectrum of $T_1$), which detects defects arising from existing TLSs.
    \item Classical flux crosstalk between nearest-neighbour (NN, such as qubit and nearby coupler) and next-nearest-neighbour (NNN, such as qubit pairs) circuits is originally maintained below 0.1\%.
\end{itemize}

The qubit spectroscopy serves as the foundation for the subsequent automated qubit bring-up process, ensuring that each qubit is initialized at its expected frequency. This procedure consists of several sequential steps, as outlined in flowchart in Fig.\ref{fig:S2A1}:

\begin{itemize}
    \item Set the flux bias of each qubit based on its spectroscopy to bring the qubit close to its desired idle frequency.
    \item Establish a suitable readout power determined from the response function referred.
    \item Configure the readout frequency to match the frequency of readout resonator when qubit is in state $\ket{0}$.
    \item Perform power Rabi oscillations to obtain an initial estimate of the 30 ns $\pi$ pulse amplitude.
    \item Fine-tune the qubit bias to the target value using Ramsey experiments, minimizing detuning errors in single-qubit gates.
    \item Recalculate the gate parameters (amplitudes) to renormalize the single-qubit gate duration into 120 ns, which reduces the frequency bandwidth of excitation pulse, thereby mitigating errors caused by classical XY crosstalk.
    \item Apply DRAG~\cite{Motzoi2009} correction (including optimizing DRAG coefficient and fine-tuning gate amplitude) to mitigate leakage out of computational subspace and control errors, thus realizing high-fidelity single-qubit gates.
    \item Optimize the readout frequency by characterizing the dispersive shift, achieving higher readout visibility and fidelity.
    \item Perform calibration for IQ readout signals to discriminate between states $\ket{0}$ and $\ket{1}$ before any probability-based readout experiment, ensuring accurate state classification.
\end{itemize}

Automated bring-up of individual qubits forms a crucial foundation for the subsequent calibration processes, such as multi-qubit energy level arrangement. This approach provides an efficient, black-box technique that enhances the overall integration of automation in our system.

\subsubsection{\label{sec:3A2}Automatic allocation of qubit idle points}

\begin{figure}[ht!]
	\centering
	\includegraphics[width=1.00\linewidth]{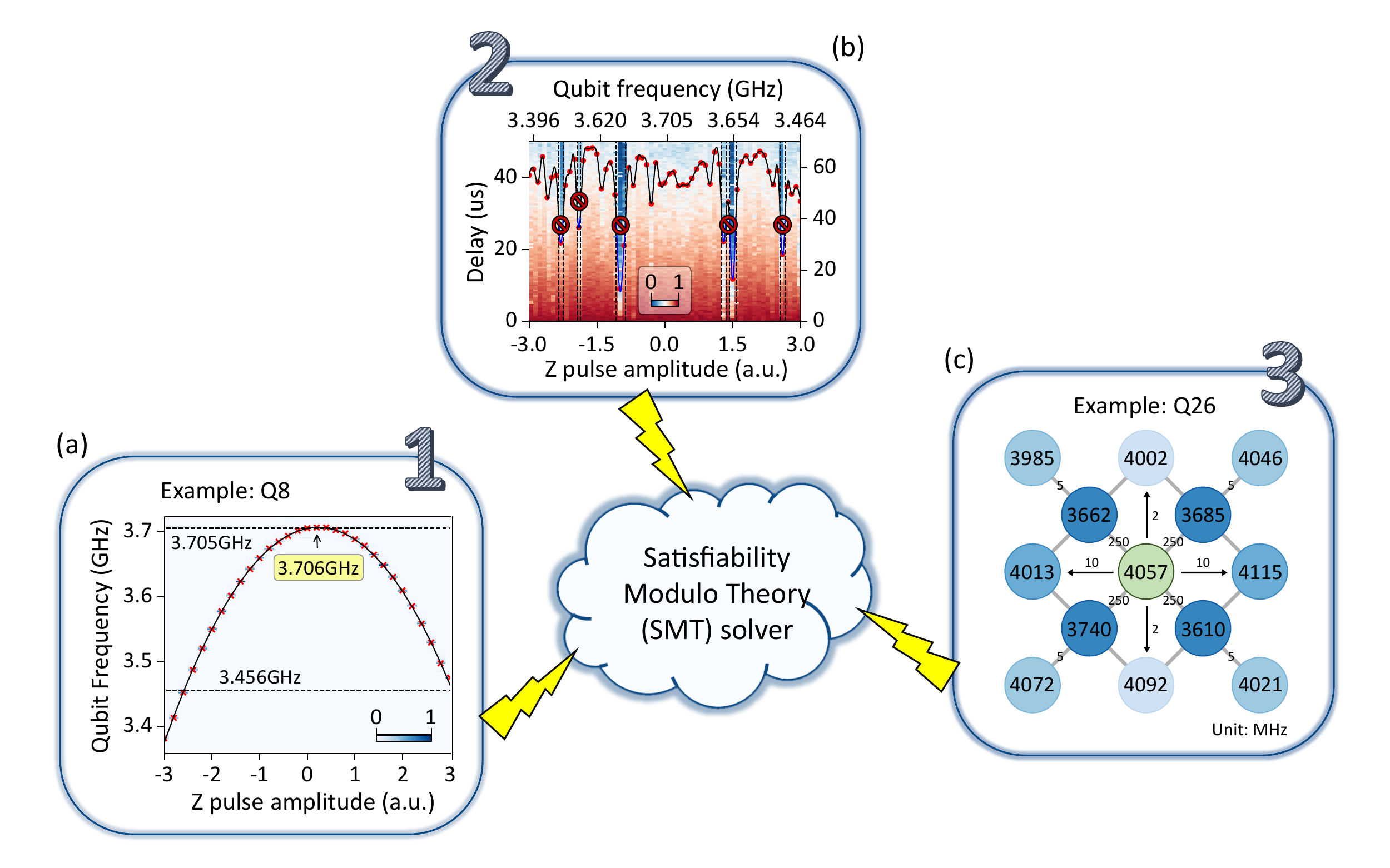}
	\caption{\textbf{Schematic of constraints for automated energy levels allocation.} (a) Constraint 1: The idle point should be neither too close nor too far from the sweet point. For example, as shown in the qubit spectroscopy of $Q_8$, an acceptable idle frequency window (1–250 MHz below the sweet point) is indicated between the two black dashed lines. (b) Constraint 2: Frequencies with poor coherence must be avoided. As shown in spectrally resolved $T_1$ for $Q_8$, the region between the black dashed lines corresponds to qubit frequencies where $T_1$ falls below an acceptable threshold. For 2D heat map, delay follows the left $y$-axis. While the red dots represent the fitted $T_1$ for experiment points, delay of which follows the right $y$-axis. The black-blue curve means the fitted $T_1$ as a function of qubit bias, where black indicates $T_1>40~\rm{\mu s}$ and blue indicates $T_1<40~\rm{\mu s}$. (c) Constraint 3: Detuning $\Delta$ must be much greater than the coupling strength $g$. A schematic of the energy level configuration around $Q_{26}$ (green) is shown. Neighboring qubits are detuned by at least 250 MHz, 10 MHz, 5 MHz, and 2 MHz, which are annotated on the links. The qubit circles are filled with color from dark blue to light blue, according to their effective coupling strength.}
	\label{fig:S2A2}
\end{figure}

Based on the fundamental information (particularly qubit spectroscopy and spectrally resolved $T_1$) obtained in \ref{sec:2A1}, we establish a series of constraints for configuring the idle energy levels of qubits:

\begin{itemize} 
    \item The idle frequency should not be set too close to or too far from the qubit sweet point. If the idle frequency is placed too near its sweet point, achieving this non-realistic target frequency might fail due to fitting errors in qubit spectroscopy. Conversely, setting idle point too far away could lead to short dephasing time ($T_2^*$). In our experiments, the idle frequency is restricted to be within 1 MHz to 250 MHz below the sweet point of the aiming qubit, as indicated between the dashed lines in Fig.~\ref{fig:S2A2}(a).
    \begin{equation}
        f^{\rm sw} - 250~{\rm MHz} < f^{\rm idle} < f^{\rm sw} - 1~{\rm MHz}
    \end{equation}
    
    \item Regions where qubits exhibit short coherence time ($T_1$) due to existing TLSs are identified and avoided, utilizing the measured spectrally resolved $T_1$, as shown in Fig.~\ref{fig:S2A2}(b).
    \begin{equation}
    f^{\rm idle} \notin \left\{f: f_{\rm TLS}\right\}
    \end{equation}
    
    \item The detuning $\Delta$ between NN and NNN qubits, should be significantly stronger than the stray coupling $g$ between them. For our devices, as illustrated in Fig.~\ref{fig:S2A2}(c), we consider directly coupled NN, two types of diagonal NNN (vertical and horizontal), and NNN connected through an intermediate qubit. The corresponding coupling strengths are approximately 4 MHz, 0.1 MHz, 1 MHz, and 0.2 MHz, respectively. Therefore accordingly, the required detunings are restricted to be greater than 250MHz (accounting for anharmonicity of transmon is around 200 MHz), 2~MHz, 10~MHz, and 5~MHz, respectively. Additionally, the detuning between any two qubits is also constrained to be at least  2MHz to avoid classical XY crosstalk.
    \begin{equation}
        \Delta \gg g,\ \Delta_{\rm idle} > \Delta_{\rm set}
    \end{equation}
\end{itemize}

This constraint optimization problem is answered by the Satisfiability Modulo Theory (SMT) solver implemented with the Python "pysmt" package~\cite{Xu2023}. With the resulting energy level allocation, the automated qubit bring-up black-box described in \ref{sec:2A1} is then applied to achieve the preparation for multi-qubit idle configuration. 

After initializing each qubit according to the SMT-optimized energy levels allocation, we configure all qubits together and perform the following series of checks and optimizations for the multi-qubit system, flowchart of which is shown in :

\begin{itemize} 
    \item Fine-tune qubit flux bias using Ramsey experiments, as residual classical Z crosstalk from other qubits is still present.
    \item Synchronize all control pulses (XY and Z signals), including those for qubits and couplers, using the automated technique described in the following \ref{sec:2A3}.
    \item For NN qubit pairs, check the ZZ coupling strength by measuring the frequency shift (via Ramsey experiments) of one qubit (target) when the other (control) is excited. Adjusting idle frequency of the coupler between the qubits can reduce this quantum ZZ crosstalk to below 0.1 MHz. 
    \item For NN qubit pairs, check the XY coupling strength by comparing the simultaneous excitation probabilities of the target qubit when the control qubit is either excited or not. Tuning the readout frequency of the coupler between the qubits can reduce the quantum XY crosstalk to within 1\%. 
    \item For NNN qubit pairs, check the XY coupling strength and determine whether the observed XY crosstalk originates from simultaneous excitation or readout. This is achieved by re-exciting (i.e., reset) the control qubit to its ground state before simultaneous readout. If the crosstalk disappears after reset, it can be then attributed to readout and can be suppressed by adjusting the qubits readout frequencies. Otherwise, this crosstalk is then induced by excitation and can only be mitigated by adjusting the qubits idle frequencies. 
    \item For qubit pairs with close-frequency readout resonators (primarily, vertical diagonal NNN pairs), the readout frequencies should be separated as much as possible to reduce classical crosstalk during simultaneous readout. Typically, the qubit readout frequency is required to far bias from its sweet point to achieve this separation.
    \item Evaluate the stability of readout fidelity for all qubits. Re-adjust the readout frequency if necessary.
\end{itemize}

Thus far, a multi-qubit idle configuration with stable readout and well-suppressed crosstalk is finally achieved. This forms a solid foundation for the subsequent calibration for simulating quantum many-body systems.

\subsubsection{\label{sec:2A3}Automatic synchronization of control pulses}

\begin{figure}[ht!]
	\centering
	\includegraphics[width=1.00\linewidth]{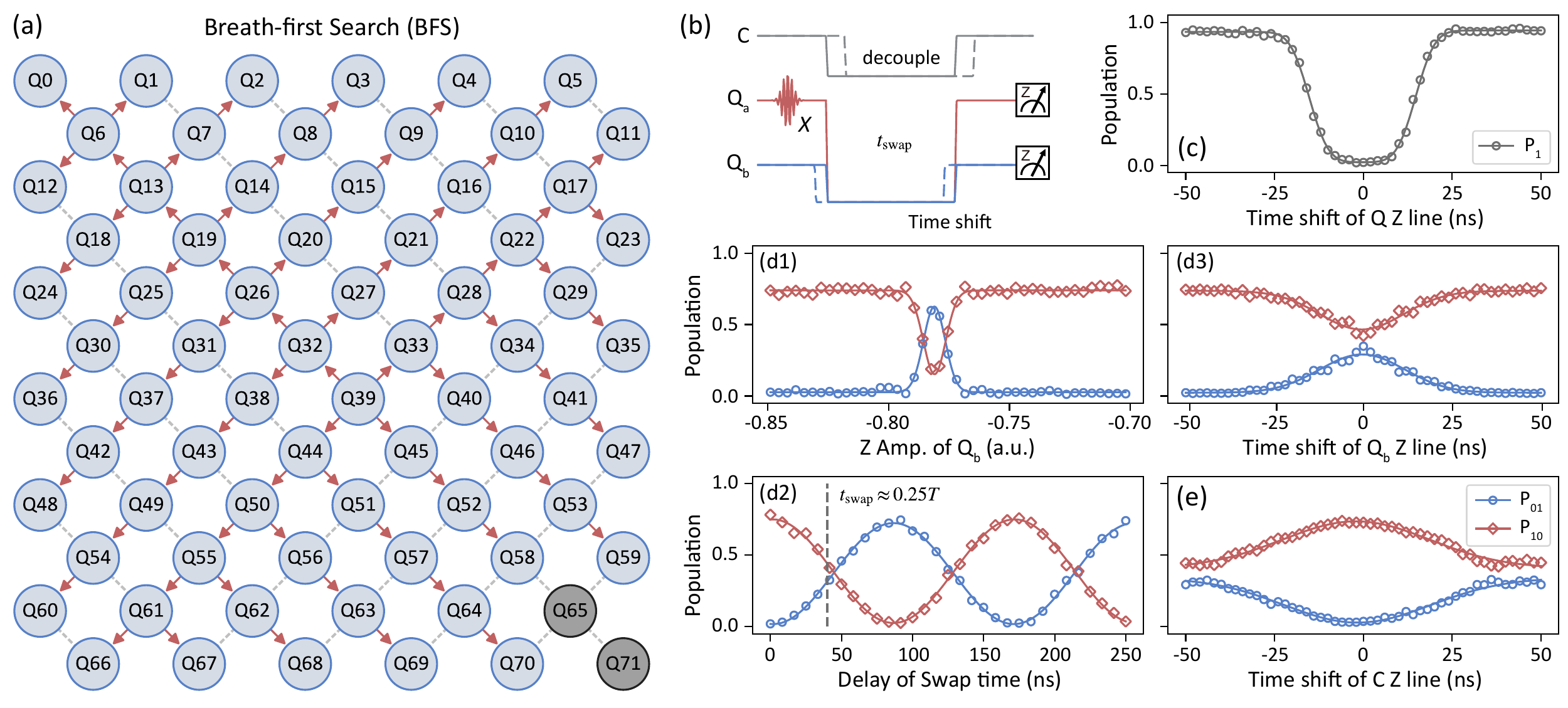}
	\caption{\textbf{Schematic and typical data for automated pulse synchronization.} (a) Schematic of the breadth-first search (BFS) algorithm starting from qubit $Q_{39}$. Red arrows indicate the propagation path, where each newly aligned qubit serves as the reference for the next ones. (b) Pulse sequences used in (d) QQswap and (e) QQswap\_C experiments. (c) Typical data for single-qubit timing calibration experiment, demonstrating synchronization of the XY and Z control lines for an individual qubit. (d) Typical data for QQswap experiment: (d1) identifying the bias point for fully coherent exchange, (d2) calibrating the exchange period $T$, and (d3) aligning the qubit pair by observing the population peak at $t_{\rm swap}\approx T/4$. (e) Typical data for QQswap\_C experiment, which synchronize Z signal of the coupler with its neighboring qubits, evidenced by maximal suppression of the swap oscillation.}
	\label{fig:S2A3}
\end{figure}

Differences in the lengths of the transmission lines used for distributing trigger signals to various electronic devices, as well as slight timing variations in the control signals transmitted through coaxial cables and microwave components, necessitate the synchronization of all control pulses. We provide an automated timing procedure (pseudo-code see Algorithm~\ref{alg:timing}) which consists of the following steps:

\begin{itemize}
    \item \text{[}Fig.\ref{fig:S2A3}(c)\text{]} Align XY and Z signals for each individual qubit (named as "single-qubit timing calibration").
    \item \text{[}Fig.\ref{fig:S2A3}(a)\text{]} Construct breadth-first search (BFS)~\cite{Xu2023} based on the chip topology, starting from its center. This algorithm not only enables rapid conduction of the synchronized timing but also generates detailed spatial coordinates for the chip structure, which can be used to create, for instance, quasi-periodic disorder patterns.
    \item \text{[}Fig.\ref{fig:S2A3}(d)\text{]} Align the timing between qubits pairs using qubit-qubit swapping experiment (named as "QQswap")~\cite{Wang2025}.
    \item \text{[}Fig.\ref{fig:S2A3}(e)\text{]} Iterate over all couplers. Based on the above qubit-qubit swapping, adjust the coupler flux bias to effectively decouple the qubits (named as "QQswap\_C"), thereby aligning the coupler Z signal with its nearby qubit.
\end{itemize}

The pulse sequences and typical data are illustrated in Fig.\ref{fig:S2A3}. In the single-qubit timing calibration experiment, the qubit is biased away from its idle point. When the XY and Z signal timing is correctly aligned, the qubit fails to excite, producing a pronounced dip in its population, as shown in Fig.\ref{fig:S2A3}(c). Furthermore, we detail the QQswap and QQswap\_C protocols, both of which begin by identifying a bias point that yields fully coherent exchange. The waveform schematic is given in Fig.\ref{fig:S2A3}(b). Initially, the bias of one qubit is finely adjusted so that its resonance (3.520~GHz in our setup) matches that of its partner, as shown in Fig.\ref{fig:S2A3}(d1). Next, the exchange period $T$ for the above resonant bias is calibrated, as depicted in Fig.\ref{fig:S2A3}(d2). A swapping time of $t_{\rm swap}\approx T/4$ is chosen as the criterion. By shifting the bias of one qubit, maximum swapping (manifested as a peak in the population of the initially unexcited qubit) occurs only when the two Z signals are synchronized (see Fig.\ref{fig:S2A3}(d3)). Following the similar exchanging working point, a coupler bias is introduced and scanned to effectively decouple the qubit pair. When the coupler Z signal is synchronized with both qubits, the swapping is then maximally suppressed, resulting in a deep minimum in the population of the unexcited qubit (Fig.\ref{fig:S2A3}(e)). Following these steps, we can establish the precise timing alignment for both qubit and coupler control lines.

\begin{algorithm}
\caption{Automatically synchronize control pulses} \label{alg:timing}
\KwData{starting qubit $Q$, chip topology $G$}
\KwResult{synchronized control pulses, including qubit XY and Z signals, and coupler Z signals}
qubits pairs $P_s(Q_a, Q_b)$ $\leftarrow$ BFS for $G$ starting from $Q$, couplers $C_s$ $\leftarrow$ $G$\;
Aligned XY and Z signal for $Q$ $\leftarrow$ Single-qubit timing calibration for $Q$\;
\For {$(Q_a, Q_b)\in P_s(Q_a, Q_b)$}
     {
        Aligned XY and Z signal for $Q_b$ $\leftarrow$ Single-qubit timing calibration for $Q_b$\;
        Timing between qubit pair $(Q_a, Q_b)$, update XY and Z signal timing for $Q_b$ $\leftarrow$ QQswap for $(Q_a, Q_b)$\;
     }
\For {$C\in C_s$}
     {
        $(Q_a, Q_b)$ $\leftarrow$ locate $C$ in $G$\;
        Update Z signal timing for $C$ $\leftarrow$ QQswap\_C for $(Q_a, Q_b)$\;
     }
\end{algorithm}

\subsection{Precise calibration for aiming Hamiltonian}

\begin{figure}[ht!]
	\centering
	\includegraphics[width=1.00\linewidth]{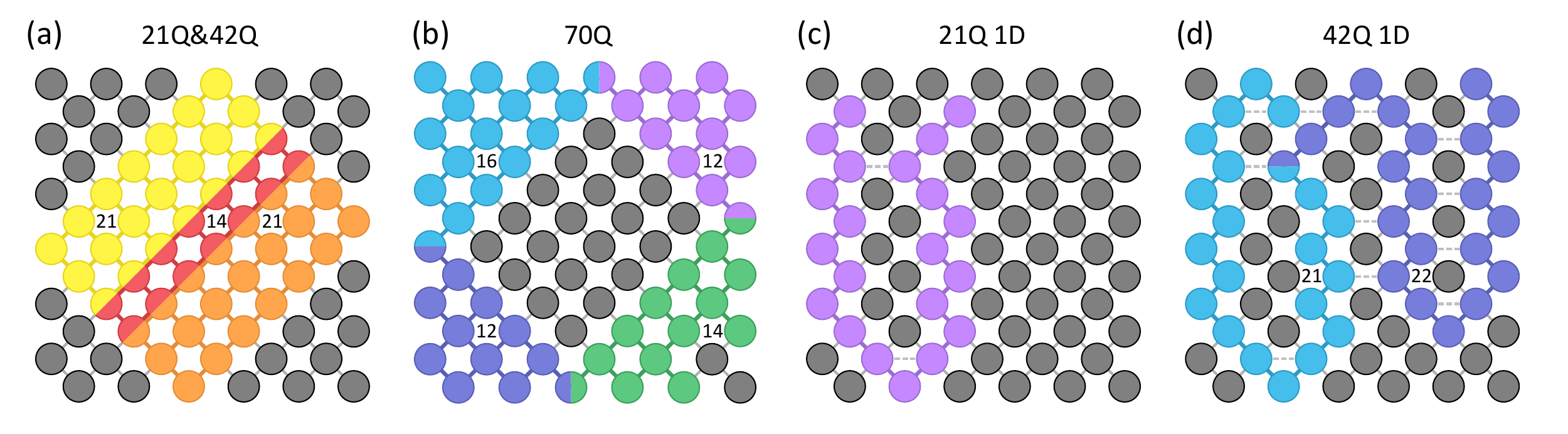}
	\caption{\textbf{Detailed block segmentation for different system sizes.} (a) Two distinct 21-qubit systems (yellow and orange) are shown, along with a two-dimensional 42-qubit system composed by combining these two blocks with an additional sandwiched 14-qubit block (red) serving as the connection. (b) Segmentation of a 70-qubit system. The central 42-qubit block (segmented identically to (a)) is surrounded by four corner subsystems containing 12, 12, 14, and 16 qubits (colored dark blue, purple, green, and light blue, respectively). (c) A one-dimensional (1D) 21-qubit chain system (purple). (d) A 42-qubit 1D chain divided into 21-qubit and 22-qubit subsystems (light blue and dark blue, respectively). The dashed gray lines in (c-d) represents the NNN couplings $J_{ij}^\mathsmaller{\rm NNN}$, which slightly non-idealize the 1D chains.}
	\label{fig:S2B1}
\end{figure}

Given the well-prepared multi-qubit configuration, we process with simulating the quantum many-body Hamiltonian after performing targeted calibrations. In this section, the calibration process is divided into two main parts: (i) \textit{calibrating bias for resonance}, which measures the specific resonant frequency using time-domain Rabi oscillations and then scans a fine spectrum of qubit transition within approximately $\pm$100 MHz around this resonant working point, and (ii) \textit{checking control bias for all circuits}, which delivers multi-qubit on-resonant experiments on different block segments (derived from 70 qubits, shown in Fig.\ref{fig:S2B1}) to verify the control parameters, including bias for qubits to its resonant working point and bias for couplers to ideal effective coupling strength, are as expected.

\subsubsection{\label{sec:2B1}Calibrate specific resonant frequency using time-domain Rabi oscillation}

Due to residue Z crosstalk and parasitic couplings, the qubit frequency may slightly deviate from its expected bare frequency when all the qubits are simultaneously biased and thus dressed~\cite{Andersen2025}. To address this issue, we introduce a calibration method that precisely determines the resonant frequency by measuring parameters under similar simultaneous biasing conditions.

\begin{figure}[ht!]
	\centering
	\includegraphics[width=1.00\linewidth]{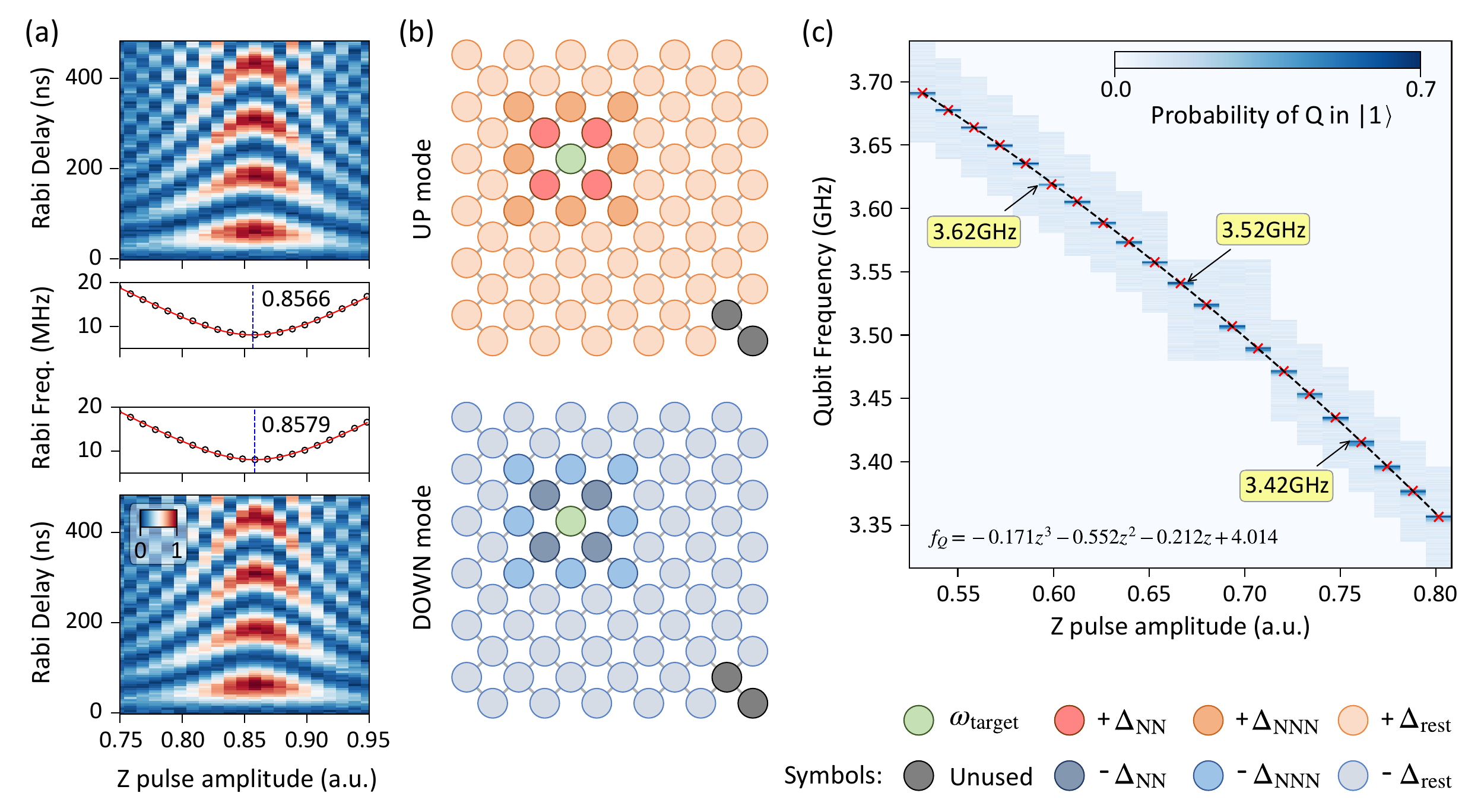}
	\caption{\textbf{Typical experimental data and configurations for calibrating on-site potentials.} (a) Time-domain Rabi oscillations measured in both UP and DOWN biasing modes. The heatmap represents the probability of qubit in $\ket{1}$. The Fourier transformation of this data shows the Rabi frequency as a function of the Z pulse amplitude. The precise Z pulse amplitude $Z_{\rm target}$ is determined as the mean amplitude corresponding to the minimum Rabi frequency observed in the two configurations. (b) Schematic configurations of UP and DOWN modes for the target qubit $Q_{26}$. The frequency detunings are set to $\Delta_{\rm NN}/2\pi$ = 80 MHz, $\Delta_{\rm NNN}/2\pi$ = 20 MHz, and $\Delta_{\rm rest}/2\pi$ = 10 MHz, respectively. In the UP mode, all (surrounding) qubits are tuned above the resonant frequency, While in the DOWN mode, they are biased below. (c) Qubit spectroscopy near the resonant frequency of 3.520~GHz over a 100~MHz range, corresponding to the maximum disorder strength. The dashed black line represents the fit using a cubic polynomial function, detail of which is shown in the inset.}
	\label{fig:S2B2}
\end{figure}

As illustrated in Fig.\ref{fig:S2B2}(b), we define two configurations for calibrating the target qubit frequency. In both cases, the NN, NNN, and remaining qubits are detuned by $\pm\Delta_{\rm NN}$, $\pm\Delta_{\rm NNN}$, and $\pm\Delta_{\rm rest}$, respectively, from the target frequency. In the first configuration (named as ``UP'' mode), all surrounding qubits are biased above the target frequency, whereas in the second configuration (named as ``DOWN'' mode), they are biased below. We then perform time-domain Rabi oscillations under both configurations, as shown in Fig.\ref{fig:S2B2}(a), and determine the precise Z pulse amplitude $Z_{\rm target}$ which brings the target qubit into resonance. This value is obtained by averaging the Z pulse amplitudes from both configurations, effectively mitigating calibration errors introduced by the interaction-induced shifts~\cite{Wang2025}.

To further enable precise control over on-site potentials (i.e., disorders), we perform the detailed qubit spectroscopy within the vicinity of the target frequency, covering the range corresponding to the maximum disorder strength (see typical data in Fig.\ref{fig:S2B2}(c)). Arbitrary on-site potentials are then realized by appropriately adjusting the Z pulse amplitude. Specifically, the required amplitude is calculated by adding the difference of the Z amplitudes between the desired disordered frequency and the resonance frequency using detailed spectroscopy to the calibrated $Z_{\rm target}$ obtained from Rabi oscillations.

\subsubsection{Block segment of multi-qubit on-resonant experiment}

We precisely calibrate the resonant working points for all 70 qubits by time-domain Rabi oscillations (detailed in \ref{sec:2B1}). Additionally, we fine-tune the effective coupling strengths to approximately 3 MHz via measuring the modulation related to coupler flux bias, and further refine accurate values through two-qubit swap spectroscopy, as discussed in \ref{sec:1D}. However, experimental imperfections inevitably lead to minor deviations. Therefore, before performing the core experiments concerning many-body delocalization, we conduct on-resonant experiments, comparing their dynamics with numerical simulations, to validate the accuracy of all these calibration parameters.

\begin{figure}[ht!]
	\centering
	\includegraphics[width=1.00\linewidth]{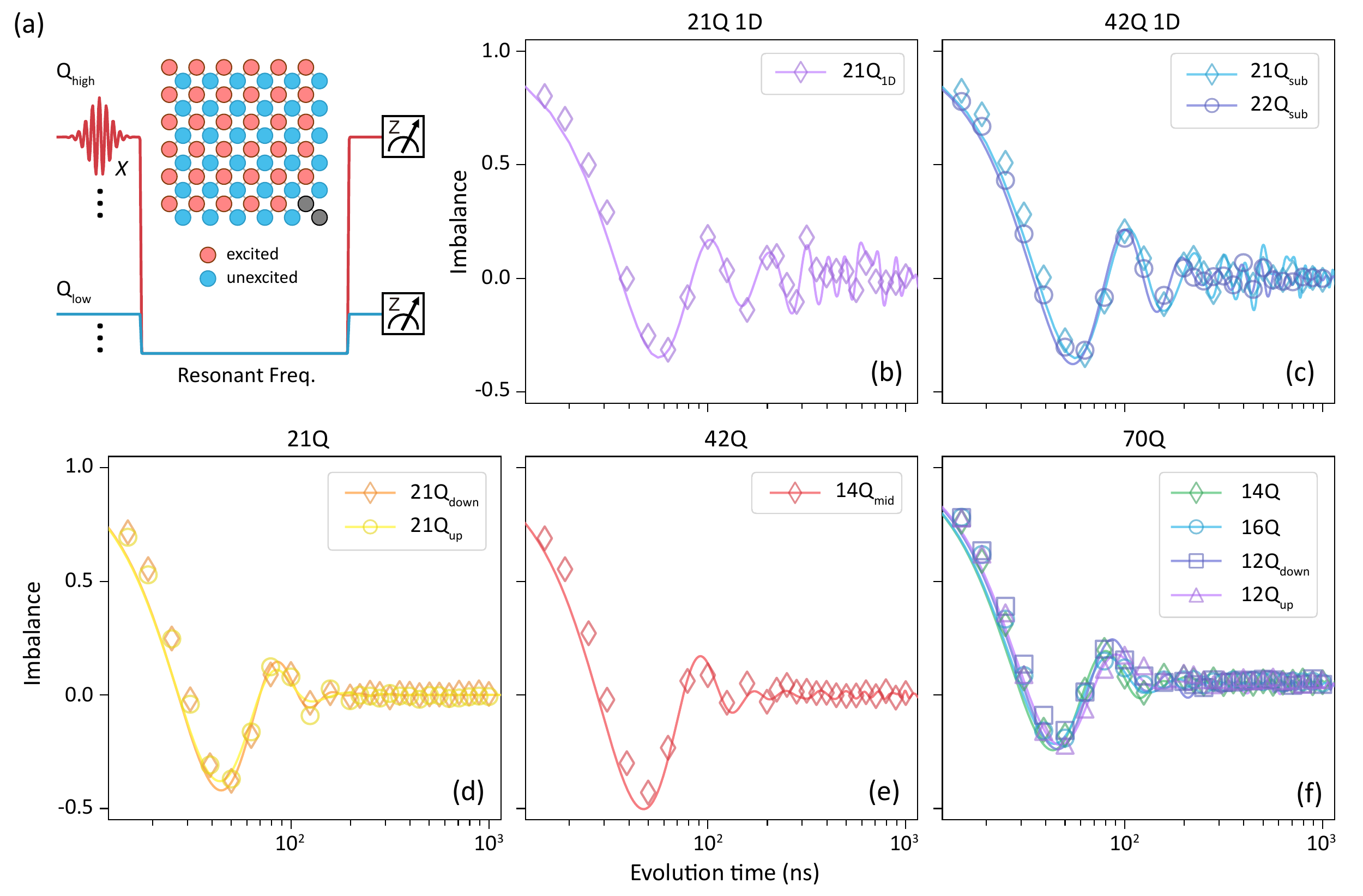}
	\caption{\textbf{Comparison between experimental and simulated thermal imbalance dynamics for different block segments.} Experimental data (diamonds) and numerical simulations (solid lines) are shown for each subsystem, with colors matching the definitions in Fig.\ref{fig:S2B1}. (a) Schematic of experimental pulse sequences for thermal experiments. The qubits with higher idle (or sweet point) frequency $Q_{\rm high}$ are prepared in $\ket{1}$, while those with lower frequency $Q_{\rm low}$ remain in $\ket{0}$. The inset specifies the position of qubits which are excited. Only qubits involved in specific subsystem are excited following the instruction of this inset, which prepares a Neel state as initialization. After excitation, all involved qubits are biased to the resonant frequency for evolution, and their final populations are measured to compute the system imbalance. (b-c) Results for one-dimensional subsystems (denoted ``1D'' and ``sub''). (d-f) Results for two-dimensional subsystems. The labels ``up'', ``down'' and ``mid'' distinguish subsystems with the same total number of qubits by their physical location on the chip.}
	\label{fig:S2B3}
\end{figure}

Accurate numerical simulations of time-dependent evolution in two-dimensional systems with more than 30 qubits are complex and computationally expensive. Moreover, thermalization is characterized by global entanglement, which limits the reliability of time dependent variational principle (TDVP) simulations based on matrix product state (MPS)~\cite{Haegeman2016} due to entanglement truncation errors. To address this, we divide the more-qubit systems into several smaller blocks, each containing no more than approximately 20 qubits. We then perform independent thermalization experiments for each block, comparing their results with exact numerical simulations calculated using the mesolve function in QuTip~\cite{Johansson2012}. For example, as illustrated in Fig.\ref{fig:S2B1}, for the 70-qubit thermalization experiment, we segment the system into seven blocks of sizes 12, 12, 14, 14, 16, 21 and 21 qubits. All individual on-resonant experiments for the 21-qubit, 42-qubit, and 70-qubit systems are performed on their respective block segments and compared with numerical simulations, as shown in Fig.\ref{fig:S2B3}. The imbalance dynamics of each block exhibit excellent agreement with the corresponding numerical predictions.

\begin{figure}[ht!]
	\centering
	\includegraphics[width=1.00\linewidth]{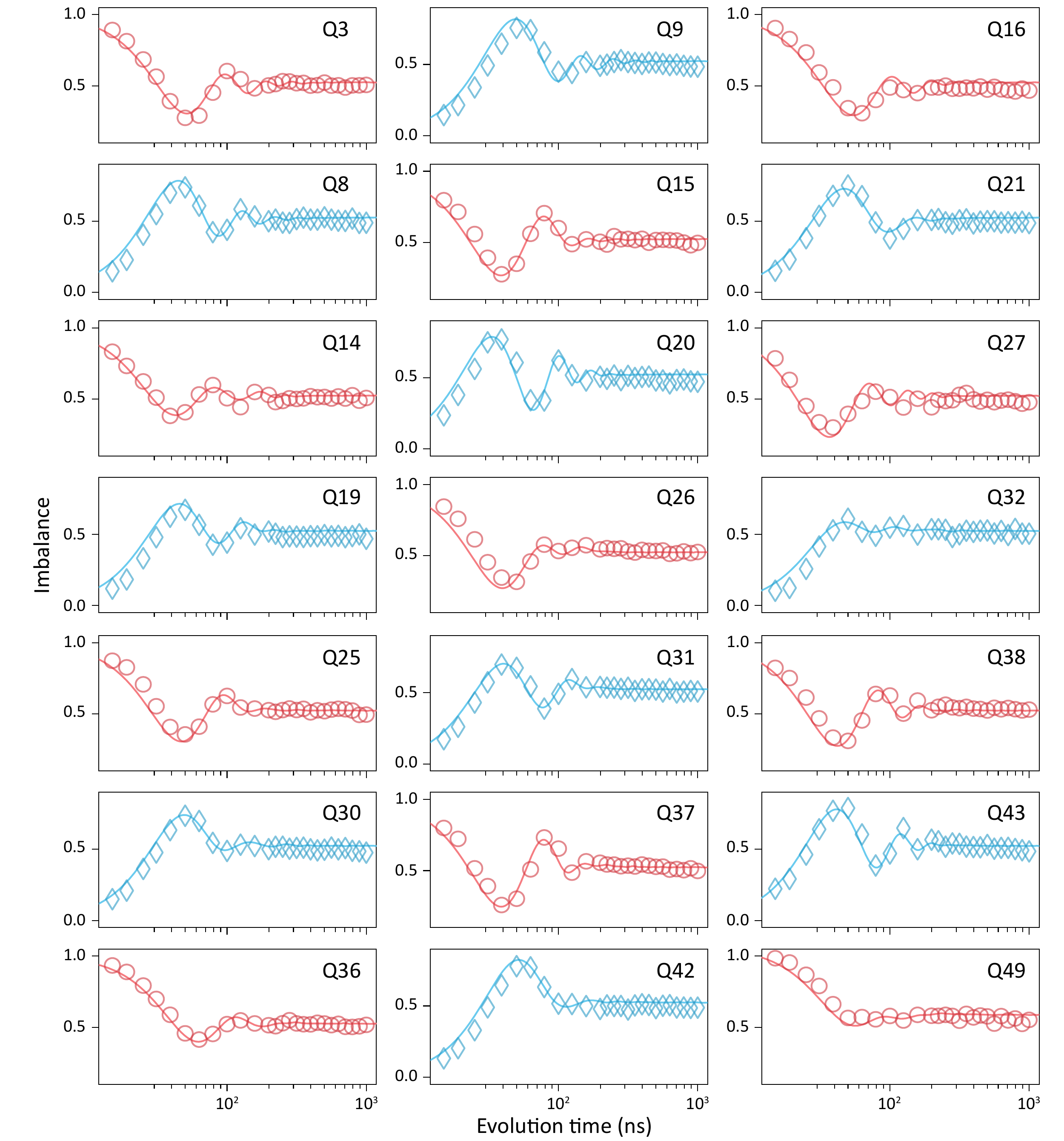}
	\caption{\textbf{Detailed comparison of single-qubit population dynamics in a 21-qubit block.} Experimental measurements (diamonds) and numerical simulations (solid lines) are shown for qubits initialized in $\ket{1}$ (red) and those remaining in $\ket{0}$ (blue). The experimental data closely match the theoretical predictions.}
	\label{fig:S2B4}
\end{figure}

Furthermore, Fig.\ref{fig:S2B4} presents a detailed comparison of the single-qubit populations for one of the 21-qubit blocks. The experimental results show excellent agreement with exact simulations, confirming that both on-resonance qubit biases and coupling strengths in our system are accurately calibrated to their true values.


\section{\label{sec:simu}Numerical methods and data}

In this section, we briefly introduce the numerical methods adopted in this work, including the polynomially filtered exact diagonalization method (POLFED) for calculating the mean level spacing $\langle r \rangle$, the Krylov subspace method for numerically simulating the quench dynamics for the systems with size $L=21$, and the time dependent variational principle (TDVP) algorithm based on the matrix product state (MPS) for simulating the dynamics for the system with size $L=42$.

\subsection{POLFED method}

The POLFED method efficiently computes eigenvalues of a Hamiltonian $\hat{H}$ close to a target energy $\sigma$. The algorithm proceeds in the following steps:

\paragraph{Step 1: Rescaling the Hamiltonian.}
First, determine the smallest and largest eigenvalues of $\hat{H}$, denoted by $e_0$ and $e_1$, respectively. The Hamiltonian is then rescaled to the interval $[-1,1]$ via:
\begin{equation}
  \hat{H}_{R} = [2\hat{H} - e_{0} - e_{1}]/ (e_{1} - e_{0}) 
\end{equation}

\paragraph{Step 2: Polynomial Spectral Transformation.}
The second step is performing the polynomial spectral transformation from $\hat{H}_{R}$ to:
\begin{equation}
  \hat{H}_{R} \rightarrow P_{\sigma}^{K}(\hat{H}_{R}) = \frac{1}{D}\sum_{n=0}^{K}c_{n}^{\sigma} T_{n}(\hat{H}_{R}),
\end{equation}
where $T_{n}(x)$ represents the $n$-th Chebyshev polynomial, $c_{n}^{\sigma} = \sqrt{4 - 3\delta_{0,n}} \cos[n \text{arccos} (\sigma) ]$, and $D$ is a normalized constant assuring that $P_{\sigma}^{K}(\sigma) = 1$. This transformation maps eigenvalues of $\hat{H}$ near $\sigma$ to the largest eigenvalues of $P_{\sigma}^{K}(\hat{H}_{R})$.

As the polynomial order $K$ increases, the filtering performance improves, but fewer eigenstates lie within the selected window. Thus, an appropriate order $K$ should be chosen such that the number of eigenvalues $\theta_i$ of $P_{\sigma}^{K}(\hat{H}_{R})$ above a fixed threshold (set to $\theta_i > p = 0.16$) equals the number of desired eigenvalues $N_{\text{ev}}$. To achieve this, we estimate the density of states $\tilde{\rho}(\sigma)$ at the target energy using the stochastic Chebyshev expansion method, and determine the appropriate value of $K$ via the bisection method.

\begin{figure}[ht!]
  \centering
  \includegraphics[width=1\linewidth]{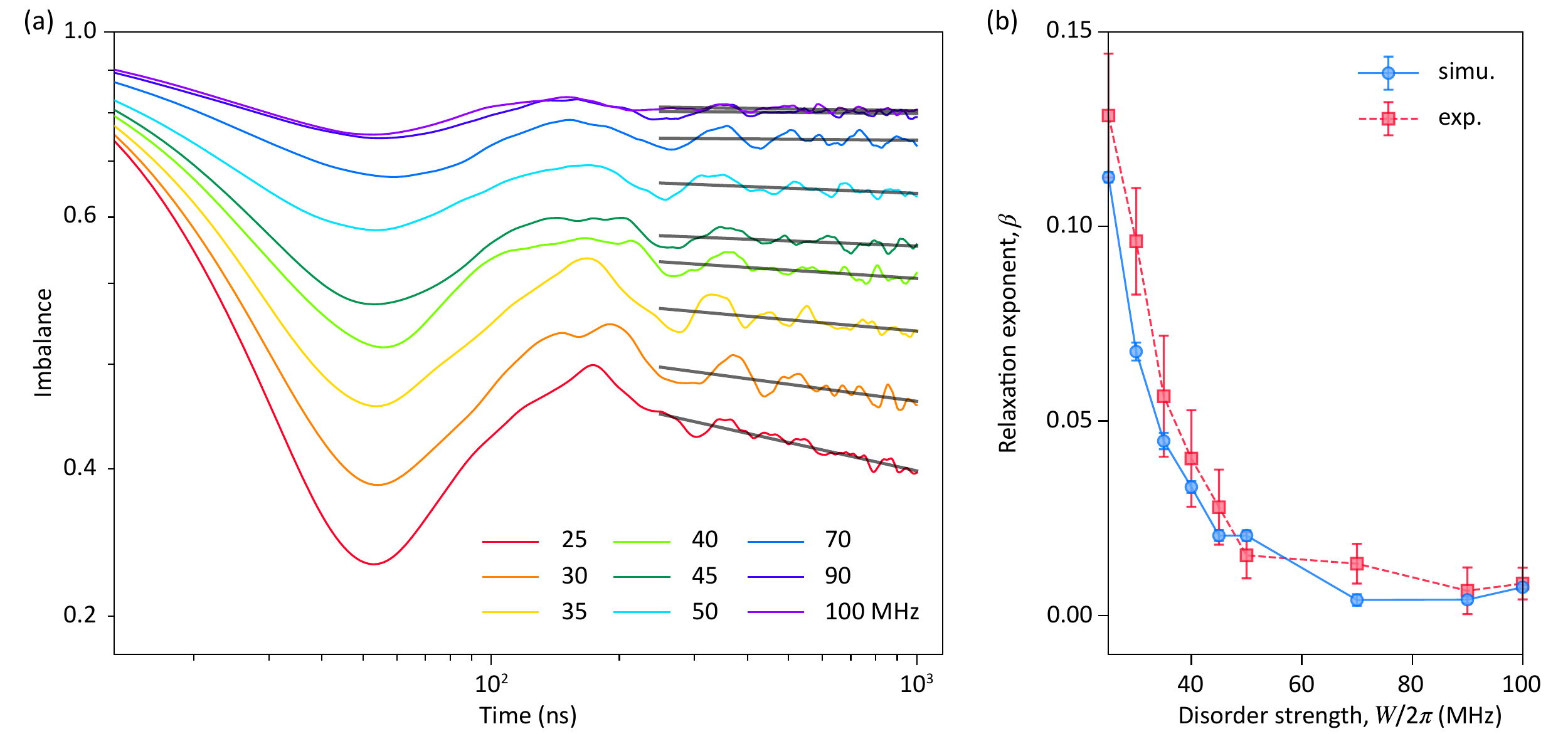}\\
  \caption{\textbf{Numerical data for 21-qubit disordered 2D systems.} (a) Numerical data of the dynamics of imbalance $I(t)$ for disordered 2D systems with the size $L=21$ and different disorder strength $W/2\pi$ ranging from $25$ MHz to $100$ MHz. The straight lines represent the power-law fitting $I(t)\propto t^{-\beta}$. (b) The relaxation exponent $\beta$ obtained from the numerical data in (a) as a function of $W/2\pi$, in comparison with the experimental data.}\label{fig:S3B1}
\end{figure}

\paragraph{Step 3: Block Lanczos Iteration.}
The third step is calculating the required largest eigenvalues of the matrix $P_{\sigma}^{K}(\hat{H}_{R})$ by using the block Lanczos iteration. 
We first generate a matrix constructed by orthonormalized random vectors $q_i$ with a dimension $\mathcal{N}\times 1$, i.e., $Q_{1} = [q_1, q_2, ..., q_s] \in\mathbb{R}^{\mathcal{N}\times s}$. The iteration proceeds as follows:
\begin{align}
&U_j = P_{\sigma}^{K}(\hat{H}_{R}) Q_j - Q_{j-1} B_j^T, \quad 
A_j = Q_j^T U_j, \\
&R_{j+1} = U_j - Q_j A_j, \quad 
Q_{j+1} B_{j+1} = R_{j+1}.
\end{align}
Here, $Q_0 = 0$ and $B_0 = 0$. After $m$ iterations, we obtain $\mathcal{Q}_m = [Q_1, Q_2, \dots, Q_m] \in \mathbb{R}^{\mathcal{N} \times ms}$,
which spans the Krylov subspace. In practice, reorthogonalization is applied after each iteration to maintain numerical stability.

\paragraph{Step 4: Extracting Eigenvalues of $\hat{H}_{R}$.}
The block tridiagonal matrix $T_m = \mathcal{Q}_m^T P_{\sigma}^{K}(\hat{H}_R) \mathcal{Q}_m $ is then diagonalized to obtain approximate eigenvectors $t_i \in \mathbb{R}^{ms}$ and $u_i = \mathcal{Q}_m t_i$. 
As the number of iterations $m$ increases, $u_i$ converge to the eigenvectors of $P_{\sigma}^{K}(\hat{H}_R)$ associated with the largest eigenvalues. Convergence can be verified by evaluating the residual norm $\| \hat{H}_{R} u_i - \epsilon_i u_i \|$. For each converged $u_i$, the eigenvalue of the rescaled Hamiltonian is given by $\epsilon_i = u_i^T \hat{H}_{R} u_i$.

\begin{figure}[ht!]
  \centering
  \includegraphics[width=1\linewidth]{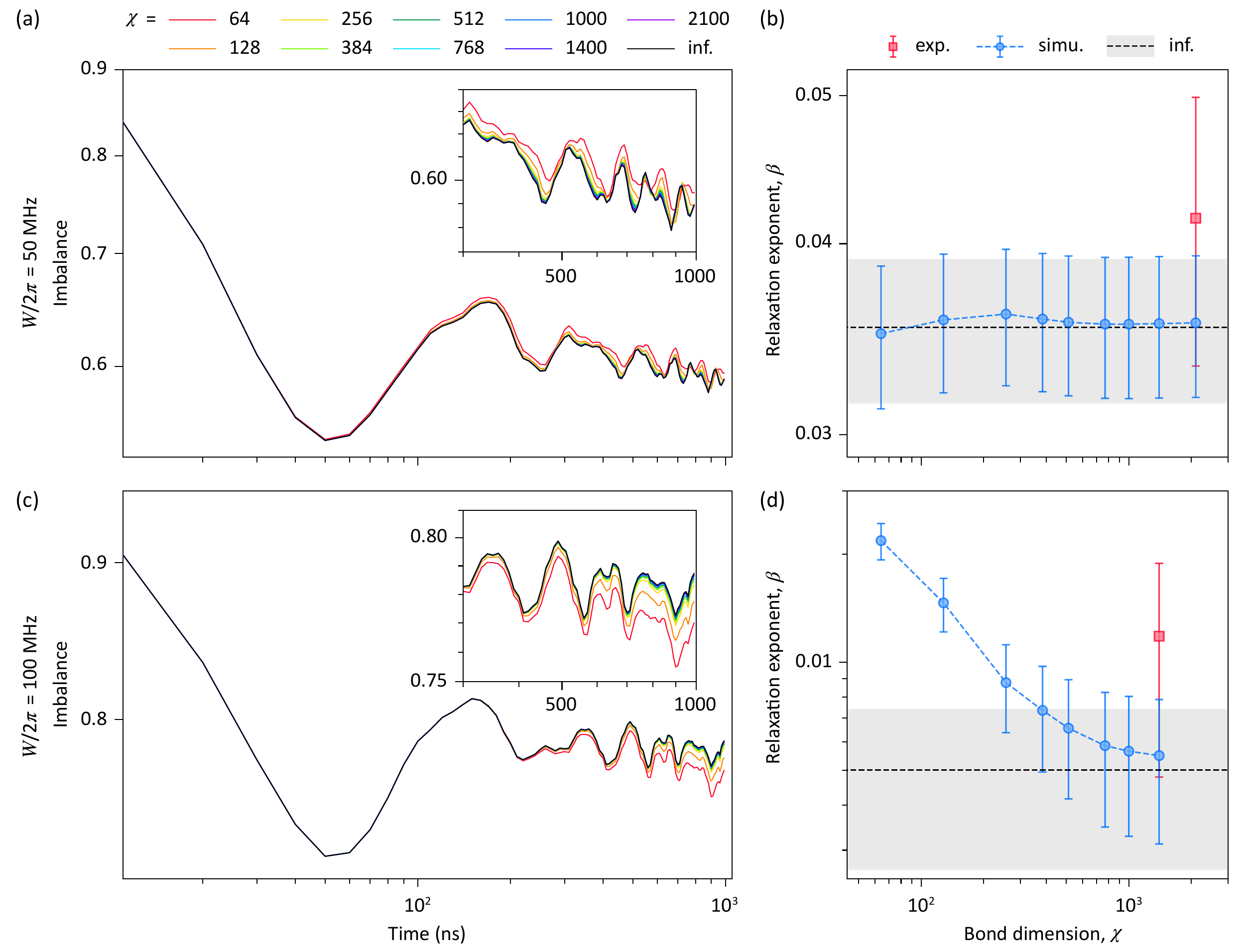}\\
  \caption{\textbf{Numerical data for 42-qubit disordered 2D systems.} (a) For disordered 2D systems with the size $L=42$ and disorder strength $W/2\pi = 50$ MHz, the dynamics of imbalance simulated by the MPS-based TDVP algorithm with different bond dimension $\chi$. The inset shows the dynamics with the time interval $t\in [250,1000]$ ns. (b) The relaxation exponent $\beta$ as a function of $\chi$ in comparison with the $\beta$ obtained from experimental data. The horizontal dashed line represents the value of $\beta$ for the estimated  imbalance with infinite bond dimension, i.e., $I(t)_{\infty}$, and the grey‑shaded band denotes the corresponding errorbar of $\beta$. (c) is similar to (a) but with the disorder strength $W/2\pi = 100$ MHz. (d) is similar to (b) but with the disorder strength $W/2\pi = 100$ MHz.}\label{fig:S3C1}
\end{figure}

\subsection{Krylov subspace method}

The unitary dynamics of a Hamiltonian $\hat{H}$, i.e., $|\psi(\Delta t )\rangle = \exp(-i\hat{H} \Delta t) |\psi_{0}\rangle $ can be exactly simulated by Krylov subspace method. The first step is constructing the Krylov subspace which is panned by the vectors $\{|\psi_{0}\rangle, \hat{H}|\psi_{0}\rangle, \hat{H}^{2}|\psi_{0}\rangle, ..., \hat{H}^{m-1}|\psi_{0}\rangle\}$ $= \{\vec{v}_{0}, \vec{v}_{1}, \vec{v}_{2}, ..., \vec{v}_{m-1} \}$, and an orthogonal matrix $K_{m} = [\vec{v}_{0}, \vec{v}_{1}, \vec{v}_{2}, ..., \vec{v}_{m-1}]$. The second step is calculating the Hamiltonian $\hat{H}$ in the $m$-dimensional Krylov subspace $H_{m} = K_{m}^{\dagger} \hat{H} K_{m}$. Finally, with an adaptively adjusted value of $m$ from $6$ to $30$ making sure that the order of errors is $10^{-14}$, the unitary dynamics can be well simulated in the Krylov subspace, i.e., $|\psi(\Delta t )\rangle = K_{m}^{\dagger} \exp(-i H_{m} \Delta t)K_{m}   |\psi_{0}\rangle $.  

In Fig.\ref{fig:S3B1}(a), we show the dynamics of the 21-qubit disordered 2D systems with different disorder strengths simulated by aforementioned Krylov subspace method. In Fig.\ref{fig:S3B1}(b), we display the relaxation exponent $\beta$ obtained by fitting the numerical data in Fig.\ref{fig:S3B1}(a). For comparison, we overlay the experimental data for the system size $L=21$ shown in Fig.3\textbf{a} of the main text, showing that the two datasets track each other closely across the full range of disorder strength $W/2\pi$.

\subsection{Matrix-product-state based time dependent variational principle}

To benchmark the experimental data with a larger system size $L=42$, we employ the time-dependent variational principle (TDVP) algorithm based on matrix product state (MPS) to simulate the dynamics. The TDVP algorithm can estimate the dynamical state $|\psi(t)\rangle $ by $\text{d}|\psi(t) \rangle / \text{d} t  = -i \mathcal{P}_{\text{MPS}} \hat{H} |\psi(t)\rangle $, with $\mathcal{P}_{\text{MPS}}$ projecting the time evolution into the variational manifold of the MPS.

In Fig.\ref{fig:S3C1}(a), we display the dynamics of imbalance for disordered 2D systems with the size $L=42$ and disorder strength $W/2\pi = 50$ MHz simulated by TDVP algorithm with different bond dimensions $\chi$. To estimate the results with $\chi\rightarrow \infty$ denoted as $I(t)_{\infty}$, we fit the data of imbalance $I(t)_{\chi}$ at the same time $t$ with $\chi$ ranging from 384 to 2100, by adopting the form of fitting function $I(t)_{\chi} = I(t)_{\infty} + a/\chi + b/\chi^{2} + c/\chi^{3}$ with the fitting parameters $\{a,b,c,I(t)_{\infty}\}$.  We also extract the relaxation exponent $\beta$ for the numerical data with different $\chi$ and $\chi\rightarrow \infty$. The results of $\beta$ are shown in Fig.\ref{fig:S3C1}(b). Similarly, for a larger disorder strength $W/2\pi = 100$ MHz, the numerical data of imbalance $I(t)$ and relaxation exponent $\beta$ are shown in Fig.\ref{fig:S3C1}(c) and (d), respectively. 

Two remarks are in order. First, in the imbalance dynamics for $W/2\pi = 50$ MHz and $100$ MHz, the results exhibit a strong dependence on the bond dimension $\chi$. Convergence of $I(t)$ is only achieved for $\chi>1000$, with reliable numerical data obtained at $\chi=1400$ [see the inset of Fig.\ref{fig:S3C1}(a) and (c)]. Second, although one might expect larger disorder to reduce entanglement (and thus computational cost), simulations at $W/2\pi = 100$ MHz remain just as demanding as those at $W/2\pi = 50$ MHz. Indeed, the relaxation exponent $\beta$ shows an even stronger $\chi$-dependence at $W/2\pi = 100$ MHz [see Fig.\ref{fig:S3C1}(d)].

\begin{figure}[ht!]
  \centering
  \includegraphics[width=1\linewidth]{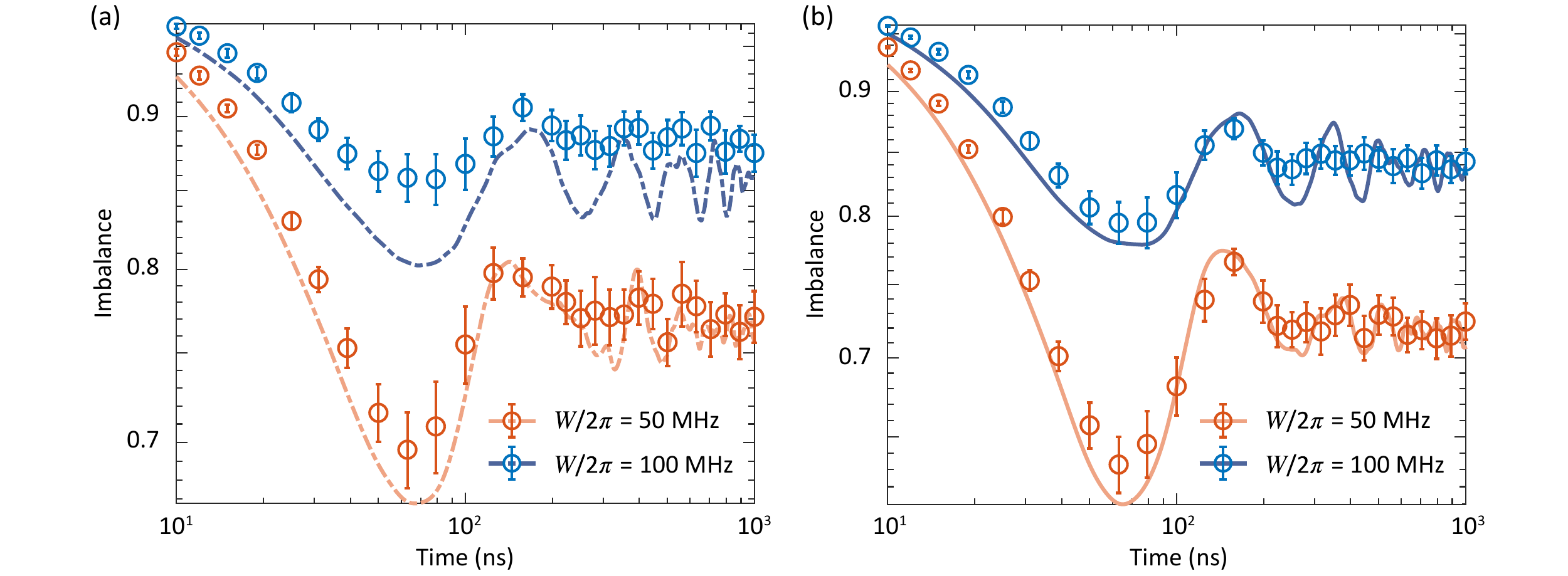}\\
  \caption{\textbf{Numerical data for disordered 1D systems.} (a) Numerical data for the dynamics of imbalance of 1D disordered systems with different disorder strengths $W$ and the system size $L=21$, in comparison with the experimental data. (b) is similar to (a) but with a larger system size $L=42$. The lines and circles represent the numerical and experimental data, respectively.}\label{fig:S3D1}
\end{figure}

\subsection{Additional numerical data for 1D disordered systems}

To benchmark the experimental data for 1D disordered systems (shown in the inset of Fig.2\textbf{b} and \textbf{c} in the main text), here we also numerically simulate the dynamics of imbalance $I(t)$ for the 1D systems with sizes $L=21$ and $42$. For $L=21$, we adopt the Krylov subspace method. For $L=42$, we employ the TDVP algorithm based on MPS with a bond dimension $\chi = 128$. In Fig.\ref{fig:S3D1} (a) and (b), we plot the numerical results of imbalance $I(t)$ for the time evolution of 1D disordered systems with different disorder strengths $W$ and the system size $L=21$ and $42$, respectively. We show that the experimental data are well consistent with the numerical results.

\section{\label{sec:data}Supplementary data}

\subsection{Dynamics of imbalance for different disorder strengths and system sizes}

\begin{figure}[ht!]
  \centering
  \includegraphics[width=1\linewidth]{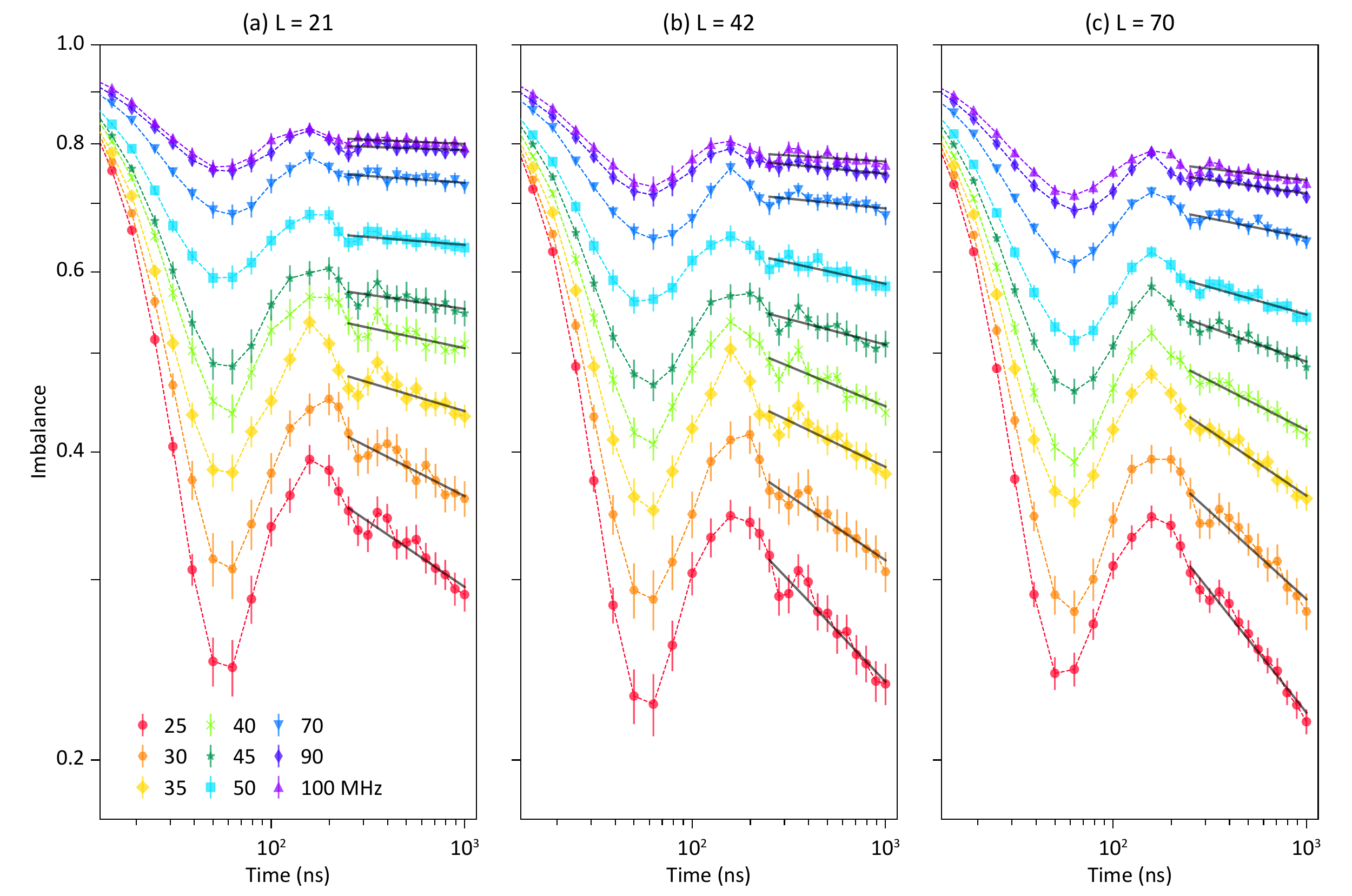}\\
  \caption{\textbf{Dynamics of imbalance of disordered 2D systems.} Experimental evolution of imbalance under varying disorder strengths (each shown with distinct colors and markers in the legend) for system sizes (a) $L=21$, (b) $L=42$ and (c) $L=70$. Solid black lines depict linear fits used to extract relaxation exponents $\beta$. Error bars represent the standard error of the statistical mean calculated over 30 disorder realizations for (a) $n=10$ and $n=11$ (60 realizations in total), (b) $n=21$, (c) $n=35$.}
  \label{fig:S4B1}
\end{figure}


%